\documentclass[iop]{emulateapj}
\usepackage{natbib,xspace,color}
\usepackage{verbatim}
\usepackage{rotating}
\usepackage{gensymb }
\usepackage{mathrsfs }  
\usepackage{amsmath}
\usepackage{amssymb }
\usepackage[backref,breaklinks,colorlinks,citecolor=blue]{hyperref}
\usepackage[all]{hypcap} 

\usepackage{url}

\newcommand{\ltir}{$L_{\rm{TIR}}$}
\shorttitle{Characterizing Dust Attenuation in Local Star-Forming Galaxies}
\shortauthors{Battisti et al.}

\begin{document}

\title{Characterizing Dust Attenuation in Local Star-Forming Galaxies: Near-Infrared Reddening and Normalization}
\author{A. J. Battisti\altaffilmark{1}, 
D. Calzetti\altaffilmark{1}
R.-R. Chary\altaffilmark{2}
}

\altaffiltext{1}{Department of Astronomy, University of Massachusetts, Amherst, MA 01003, USA; abattist@astro.umass.edu}
\altaffiltext{2}{MS314-6, U.S. Planck Data Center, California Institute of Technology, 1200 East California Boulevard, Pasadena, CA 91125, USA}

\begin{abstract}
We characterize the near-infrared (NIR) dust attenuation for a sample of $\sim$5500 local ($z\lesssim0.1$) star-forming galaxies and obtain an estimate of their average total-to-selective attenuation $k(\lambda)$. We utilize data from the United Kingdom Infrared Telescope (UKIRT) and the Two Micron All-Sky Survey (2MASS), which is combined with previously measured UV-optical data for these galaxies. The average attenuation curve is slightly lower in the far-UV than local starburst galaxies, by roughly $15\%$, but appears similar at longer wavelengths with a total-to-selective normalization at $V$-band of $R_V=3.67\substack{+0.44 \\ -0.35}$. Under the assumption of energy balance, the total attenuated energy inferred from this curve is found to be broadly consistent with the observed infrared dust emission (\ltir) in a small sample of local galaxies for which far-IR measurements are available. However, the significant scatter in this quantity among the sample may reflect large variations in the attenuation properties of individual galaxies. We also derive the attenuation curve for sub-populations of the main sample, separated according to mean stellar population age (via $D_n4000$), specific star formation rate, stellar mass, and metallicity, and find that they show only tentative trends with low significance, at least over the range which is probed by our sample. These results indicate that a single curve is reasonable for applications seeking to broadly characterize large samples of galaxies in the local Universe, while applications to individual galaxies would yield large uncertainties and is not recommended.
\end{abstract} 

\section{Introduction}
The spectral energy distribution (SED) of a galaxy is a powerful tool with which numerous physical properties, such as the stellar mass, star formation rate (SFR), and star formation history (SFH), can be inferred if sufficient wavelength coverage is available \citep[see review by][]{conroy13}. From a detailed census of these properties as a function of cosmic time, we are able to develop an understanding of galaxy formation and evolution \citep[e.g., ][]{madau&dickinson14,somerville&dave15}. In this respect, understanding the impact of dust grains on the SEDs of galaxies is vital for achieving a more accurate and detailed understanding in future studies. This is especially true at intermediate redshifts ($z\sim1-3$) where the SEDs of galaxies appear more heavily attenuated by dust than in the local Universe \citep{lefloch05,magnelli09,elbaz11,murphy11a,reddy12}. 

The best way to characterize the wavelength-dependent effects of dust is to measure extinction\footnote{We define extinction to be the combination of absorption and scattering of light out of the line of sight by dust (no dependence on geometry).} along sightlines to individual stars. In this manner, extinction curves have been developed for the Milky Way \citep[MW; e.g.,][]{cardelli89,fitzpatrick99}, Magellanic Clouds \citep[e.g., ][]{gordon03}, and the Andromeda galaxy \citep[M31; e.g.,][]{bianchi96,clayton15}. Common among these curves, which are usually expressed in terms of the total-to-selective extinction $k(\lambda)\equiv A_\lambda /E(B-V)$, is that the highest extinction occurs in the ultraviolet (UV) and it decreases with increasing wavelengths out to the infrared (IR). Such a behavior is directly related to the size distribution and composition of dust grains \citep[see review by][]{draine03}.

In contrast to the previously mentioned extinction studies, characterizing the wavelength-dependent nature of dust in distant galaxies has proven to be a difficult task. The reason for this is twofold; 1) a reliance on using unresolved stellar populations composed of many stars over a range of ages and spectral type, for which the underlying SED is a complex function of the SFH and the stellar initial mass function (IMF), and 2) extended light emitting regions can have a range of possible geometries with respect to dust, which strongly influences the underlying behavior of the dust attenuation \citep[see review by][and references therein]{calzetti01}. The second effect is the reason why it is important to distinguish between extinction and attenuation\footnote{We define attenuation to be a combination of extinction and scattering of light into the line of sight by dust (strong dependence on geometry).}, where attenuation is what occurs in extended emitting regions, including distant galaxies. These difficulties can be overcome by using large samples of galaxies with similar SFHs, which limits the impact of the first issue. This was first performed by \citet{calzetti94,calzetti00} using a sample of 39 local starburst (SB) galaxies and has since been used in other studies \citep[e.g.,][]{wild11,reddy15}. 

The SB attenuation curve of \citet{calzetti00} has been extensively utilized by the community to correct for effects of dust in galaxies, often for cases with different properties than SB galaxies and/or at higher redshift where it might not be appropriate. In fact, the results of \citet{reddy15} using a sample of 224 star-forming galaxies (SFGs) at $z\sim2$, suggest that the attenuation curve of galaxies at higher redshifts has a lower normalization than the local SB curve. This finding clearly illustrates the need to develop an understanding of the factors which influence the behavior of dust attenuation in order to accurately correct for it. The best manner in which to do this is to utilize much larger samples of galaxies to examine the behavior of dust attenuation as a function of various physical properties in those galaxies.

Recently, we began the task of characterizing the variation in the selective attenuation curve in the UV-optical wavelength range ($1250-8200$~\AA) using a sample of 9813 local SFGs \citep{battisti16}. That study found that the \emph{average} selective attenuation in the UV and optical for SFGs does not change significantly when examining sub-populations separated according stellar mass, specific star formation rate (sSFR), or mean stellar population age, although there is noticeable scatter in the attenuation among individual galaxies. In this study, we extend this analysis to the NIR ($0.82-2.1$~\micron) in order to determine the normalization of the attenuation curve derived in Battisti et al. and to determine if this normalization shows variation. 

Throughout this work we adopt a $\Lambda$-CDM concordance cosmological model, $H_0=70$~km/s/Mpc, $\Omega_M=0.3$, $\Omega_{vac}=0.7$. To avoid confusion, we make explicit distinction between the color excess of the stellar continuum $E(B-V)_{\mathrm{star}}$, which traces the reddening of the bulk of the galaxy stellar population, and the color excess seen in the nebular gas emission $E(B-V)_{\mathrm{gas}}$, which traces the reddening of the ionized gas around massive stars located within HII regions. In principle these two parameters need not be related because the gas and stars can follow different attenuation curves (see \S~\ref{method_tau}), but they have been found to be correlated in SB galaxies and star-forming regions of local galaxies with $\langle E(B-V)_{\mathrm{star}}\rangle\sim0.5\langle E(B-V)_{\mathrm{gas}}\rangle$ \citep{calzetti00,kreckel13,battisti16}. We define the total infrared luminosity as $L_{\rm{TIR}}=\int_{3\mu\rm{m}}^{1100\mu\rm{m}} L_\nu d\nu$. 

\section{Data and Measurements}
\subsection{Sample Selection}\label{data}
The parent sample of galaxies used in this study are the same as in \citet{battisti16}, and we refer the reader to that paper for a detailed description. In brief, we select galaxies from the \textit{Galaxy Evolution Explorer} \citep[\textit{GALEX}; ][]{martin05,morrissey07} data release 6/7 catalogs of \citet{bianchi14} with FUV ($1344-1786$~\AA) and NUV ($1771-2831$~\AA) detections of $S/N>5$. These sources are cross-matched with spectroscopically observed galaxies in the Sloan Digital Sky Survey (SDSS) data release 7 \citep[DR7;][]{abazajian09}, as we utilize emission line diagnostic to remove sources with a significant active galactic nucleus (AGN) contribution from the sample and to determine the dust content (through the Balmer decrement). Using this selection criteria, together with the requirement of $z\le0.105$ to avoid the FUV band being affected by absorption features, provides a sample of 9813 SFGs which acts as the parent sample for this study.

We perform a search for sources with NIR detections in the UKIRT (United Kingdom Infrared Telescope) Infrared Deep Sky Survey \citep[UKIDSS; ][]{hewett06}. UKIDSS utilized the UKIRT Wide Field Camera \citep[WFCAM; ][]{casali07}, with the calibration described in \citet{hodgkin09}, and consists of five surveys that cover a range of areas and depths. For the purpose of this study we are only interested in using the Large Area Survey \citep[LAS; ][]{lawrence07} because it lies within the SDSS footprint. The LAS observed a 4000~deg$^2$ region area in $Y$, $J$, $H$, and $K$ ($\lambda_{\mathrm{eff}}=1.0305$, 1.2483, 1.6313, and $2.2010~\micron$) to depths of $m_{\mathrm{Vega}}=20.3$, 19.5, 18.6, and 18.2 mag, respectively (5$\sigma$ detection). We cross-matched sources in the Data Release 10 (DR10) of the LAS (\texttt{lasSource}) using the SDSS coordinates with a search radius of $2\arcsec$ in the WFCAM Science Archive\footnote{\url{http://wsa.roe.ac.uk/}} \citep[WSA; ][]{hambly08}. We exclude galaxies for which the photometry is labeled as having been ‘de-blended’ because the flux in such objects has been shown to be overestimated \citep{hill10}. We select the $5.7\arcsec$ diameter aperture magnitude for the $YJHK$ bands (\texttt{y,j\_1,h,kAperMag6}) and find 3018 SFGs from the parent sample with $S/N>3$ in all four bands.

Given that the UKIDSS LAS does not cover the entire SDSS footprint, we also perform a search for sources with NIR detections in the Two Micron All-Sky Survey \citep[2MASS; ][]{skrutskie06} to identify additional sources for better statistics. The 2MASS observed the entire sky in $J$, $H$, and $K_S$ ($\lambda_{\mathrm{eff}}=1.235$, 1.662, and $2.159~\micron$) to depths of $m_{\mathrm{Vega}}=15.8$, 15.1, and 14.3 mag, respectively (10$\sigma$ detection). As the majority of galaxies in this study have a small angular extent on the sky relative to the PSF of 2MASS ($\sim$2.8$\arcsec$), almost all cases with detections are contained within the 2MASS Point Source Catalog \citep[PSC; ][]{cutri03}. This cross-matching is performed using the NASA/IPAC Infrared Science Archive (IRSA) GATOR service\footnote{\url{http://irsa.ipac.caltech.edu/Missions/2mass.html}} with a search radius of $3\arcsec$ from the SDSS coordinates. We select the standard 2MASS aperture magnitude for the $JHK_S$ bands (\texttt{j,h,k\_m\_stdap}), which have a diameter of $8\arcsec$. To mitigate the issue of sight-line contamination for this relatively large aperture, we exclude cases where multiple components are detected within $~$5$\arcsec$ (\texttt{bl\_flg}$>$1). Because the PSC is intended primarily for stellar objects, the aperture magnitudes provided are corrected to total values assuming a point-like source. For the purposes of this study we want to remove this correction and utilize fixed aperture values. We remove the aperture correction using the information in the 2MASS Atlas Image Info available through IRSA, which specifies the correction applied to sources within a given image frame. We identify the appropriate image frame using the \texttt{scan}, \texttt{coadd}, and \texttt{scan\_key} number provided in the PSC. We have checked that the aperture photometry performed in this manner are in agreement with the aperture data from a subset of the sample also available from the 2MASS Extended Source Catalog \citep[XSC; ][]{jarrett00}, where no aperture corrections were applied. For the 2MASS data, we find 3560 SFGs with $S/N>3$ in all three bands.

\begin{figure}
\begin{center}
\includegraphics[width=3.5in]{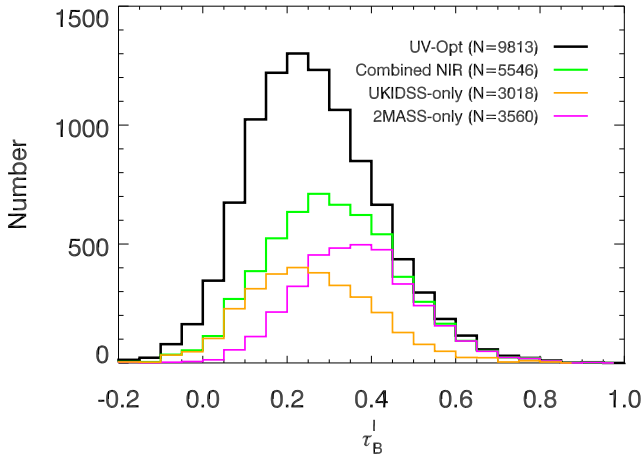}
\end{center}
\caption{Distribution of Balmer optical depths, $\tau_B^l$, a probe of dust content in galaxies, for the UKIDSS LAS (orange line) and the 2MASS (magenta line) samples compared to the parent sample selected in the UV-optical \citep[black line; ][]{battisti16}. The deep LAS provides a similar distribution as the parent sample but because it only covers a portion of the SDSS footprint, it provides poor statistics on galaxies with large $\tau_B^l$, which tend to be more massive galaxies. The 2MASS is relatively shallow but covers the entire SDSS footprint and provides excellent sampling of galaxies with large $\tau_B^l$. Combining unique galaxies in the two surveys (green line) increases our NIR sample size and provides a wider range of $\tau_B^l$ for use in deriving the attenuation curve (\S~\ref{derive_curve}). \label{fig:tau_hist}}
\end{figure}

For our analysis we combine the data from the LAS and 2MASS because they offer complementary views of the parent UV-optical sample. The LAS provides a good census of the more common faint galaxy population but lacks rarer, bright objects, which are abundant in the  shallower but full-sky 2MASS. This wider flux density sampling is important because brighter (more massive) galaxies tend to experience more dust attenuation \citep[e.g.,][]{wang&heckman96,hopkins01,garn&best10,reddy15,battisti16} and probing a wider range in dust content provides more accuracy in characterizing the dust attenuation. We demonstrate the difference in dust content, as probed by the Balmer optical depth ($\tau_B^l$, see \S~\ref{method}), for the different samples in Figure~\ref{fig:tau_hist}. Combining the two samples provides NIR data for 5546 unique SFGs (1032 galaxies are observed in both surveys) and gives a wider range in $\tau_B^l$ than is possible with each survey individually. When data are available in both surveys for a galaxy we utilize the deeper LAS data because it has a better $S/N$. 

All measurements of galaxy properties utilized in this work are from the Max Planck Institute for Astrophysics and Johns Hopkins University (MPA/JHU) group\footnote{\url{http://www.mpa-garching.mpg.de/SDSS/DR7/}} and correspond to the 3\arcsec\ SDSS fiber, which is typically centered on the nuclear region and represent a fraction of the total galaxy. We adopt the uncertainty values listed in \citet{juneau14} for the emission lines, which are updated values for the DR7 dataset. The stellar masses are based on fits to the photometric data following the methodology of \citet{kauffmann03a} and \citet{salim07}. The SFRs are based on the method presented in \citet{brinchmann04}. The gas phase metallicities are estimated using \citet{charlot&longhetti01} models as outlined in \citet{tremonti04}. All photometry and spectroscopy has been corrected for foreground Milky Way extinction using the \textit{GALEX} provided $E(B-V)_{\mathrm{MW}}$ with the extinction curve of \citet{fitzpatrick99}. The fiber regions of the 5546 SFGs in this sample span the following range in properties: $6.31<\log[M_* (M_{\odot})]<10.67$ (median=$9.34$), $-3.43<\log[SFR (M_{\odot}\mathrm{yr}^{-1})]<1.60$ (median=$-0.36$), and $7.91<12+\log(O/H)<9.37$ (median=$8.94$).

\subsection{Optical-NIR Aperture Matching}
The aperture photometry obtained from the LAS and 2MASS source catalogs are larger than the $4.5\arcsec$ aperture used in \citet{battisti16}, and require a correction to match the previous data for deriving the underlying attenuation curve in the NIR. We perform the corrections using the annular light profile catalog (\texttt{PhotoProfile}) from the SDSS, which are obtained through the CasJobs website\footnote{\url{http://skyserver.sdss.org/casjobs/}}. This catalog provides annular aperture photometry at the radii of 0.22, 0.67, 1.03, 1.75, 2.97, 4.58, 7.36, and $11.42\arcsec$ (as well as larger values, but these are not utilized), which we convert into cumulative flux densities. We then linearly interpolate these flux densities to 4.5, 5.7, and $8.0\arcsec$. We normalize the UKIDSS and 2MASS photometry such that the ratio between the NIR and the SDSS $z$-band flux density at the catalog aperture is preserved for $4.5\arcsec$ (e.g., UKIDSS: $f_J(4.5\arcsec)=f_z(4.5\arcsec)[f_J(5.7\arcsec)/f_z(5.7\arcsec)]$; 2MASS: $f_J(4.5\arcsec)=f_z(4.5\arcsec)[f_J(8\arcsec)/f_z(8\arcsec)]$). 

As a check on the consistency between the datasets, we compare the aperture corrected photometry for the 1032 galaxies observed in both surveys. The effective wavelengths of the UKIDSS and 2MASS filters are slightly different and to account for this we linearly interpolate flux densities, in log($F_\lambda$)-log($\lambda$) space (this region behaves as $F_{\nu}\propto\nu^{\alpha}$), to match at the effective wavelength of the 2MASS filters. The behavior of the NIR spectral region is relatively smooth such that interpolating values for such small wavelength offsets are sufficient. We find that the fractional difference between the aperture corrected flux densities in the different bands have symmetric distributions with average values of $|\mu|<3\%$ with $1\sigma<20\%$, indicating good general agreement (see Appendix~\ref{NIR_compare} for details). The large dispersion between the samples is driven by the large uncertainty for many of the faint 2MASS sources that overlap with UKIDSS, where the majority of these 2MASS sources have $5>S/N>10$.

\section{Methodology for Characterizing Attenuation}\label{method}
\subsection{Balmer Optical Depth}\label{method_tau}
The dust attenuation in each galaxy is quantified through the Balmer decrement, $F(\mathrm{H}\alpha)/F(\mathrm{H}\beta)$, where H$\alpha$ and H$\beta$ are located at 6562.8~\AA\ and 4861.4~\AA, respectively. In the absence of an AGN, these lines arise primarily from HII regions and their flux ratio is set by quantum mechanics and only affected by environmental properties (electron temperature, $T_{\mathrm{e}}$, and density, $n_{\mathrm{e}}$) at the $\sim5-10\%$ level \citep{osterbrock06}. This ratio is also insensitive to the underlying stellar population and IMF \citep{calzetti01}. Therefore, galaxies without AGN for which the Balmer decrement deviates from the intrinsic value are experiencing reddening from dust. 

Following \citet{calzetti94}, we define the Balmer optical depth as
\begin{equation}\label{eq:tau}
\tau_B^l = \tau_{\mathrm{H}\beta} - \tau_{\mathrm{H}\alpha} = \ln \left(\frac{F(\mathrm{H}\alpha)/F(\mathrm{H}\beta)}{2.86}\right)\,,
\end{equation}
where $F($H$\alpha)$ and $F($H$\beta)$ are the flux of the nebular emission lines, and the value of 2.86 comes from the theoretical value expected for the unreddened ratio of $F(\mathrm{H}\alpha)/F(\mathrm{H}\beta)$ undergoing Case B recombination with $T_{\mathrm{e}}=10^4$~K and $n_{\mathrm{e}}=100$~cm$^{-3}$ \citep{osterbrock89,osterbrock06}. The superscript \textit{l} is used to emphasize that this quantity is coming from emission lines and should be distinguished from optical depths associated with the stellar continuum. 

If one assumes knowledge of the total-to-selective extinction, $k(\lambda)\equiv A_\lambda /E(B-V)$, then $\tau_B^l$ can be directly related to the color excess of the nebular gas, $E(B-V)_{\mathrm{gas}}$, through
\begin{equation}\label{eq:EBV_gas}
E(B-V)_{\mathrm{gas}}=\frac{A(\mathrm{H}\beta)-A(\mathrm{H}\alpha)}{k(\mathrm{H}\beta)-k(\mathrm{H}\alpha)}=\frac{1.086\tau_B^l}{k(\mathrm{H}\beta)-k(\mathrm{H}\alpha)} \,,
\end{equation}
where $A(\lambda)$ is the total extinction at a given wavelength. If the extinction in other galaxies at these wavelengths were to be identical to the MW, which is unlikely to always be the case, then we can use $k(\mathrm{H}\alpha)-k(\mathrm{H}\beta)=1.257$ \citep{fitzpatrick99}.

\section{NIR Dust Attenuation Curve}\label{derive_curve}
\subsection{Galaxy Templates}\label{templates}
To empirically determine an attenuation curve, we need to compare the SEDs of galaxies with differing amounts of dust attenuation. This task is made complicated by the fact that galaxies are composed of a collection of many stellar populations of different ages, and their SED can vary significantly from one galaxy to another depending on the SFH. Related to this issue is the well known age/dust degeneracy, for which the reddening of a young, dusty stellar population can be similar to that of an old, dust-free population \citep[e.g., ][]{witt92,gordon97}. This degeneracy can be resolved by selecting samples of galaxies with roughly similar SFHs and stellar populations, which can be selected in numerous ways (e.g., emission lines, the strength of the 4000~\AA\ break feature \citep[$D_n4000$;][]{kauffmann03a}, specific star formation rate). The issue can be further mitigated by using an ensemble of many similar galaxies to construct templates with ``average'' stellar populations as a function of attenuation \citep{calzetti94,reddy15}. We stress that in order for such templates to provide meaningful information about the dust attenuation, it is necessary that the average age of the stellar populations in these galaxies be independent of the relative attenuation. 

For this sample of galaxies, a linear relationship is observed between $\beta_{\rm{GLX}}$ and $\tau_B^l$, although with a very large dispersion \citep{battisti16}. Similar relationships are observed for local starburst galaxies \citep{calzetti94} and also in SFGs at $z\sim2$ \citep{reddy15}. In principle, the behavior of this relationship can be used to constrain the geometric distribution of the stars and dust \citep[see][]{calzetti01}. For the purpose of this analysis, we assume a foreground-like dust component is dominating the attenuation, which is consistent with the linear relationship between these quantities, and provides us with a straightforward manner to derive the effective attenuation curve. However, we stress that other geometries are not ruled out given the significant scatter in the observed relationship. For example, the more realistic two-component model presented by \citet{charlot&fall00} \citep[see also][]{wild11} also predict near-linear relationships between $\beta_{\rm{GLX}}$ and $\tau_B^l$. An implementation of this geometry would be less straightforward and would not change the quantitative results or practical implications of our analysis. We also emphasize that the size-scale being considered for these measurements is important. In a study of M83, \citet{liu13} found that on small scales ($\sim$10~pc) the star/dust geometry shows wide variation, but when averaging on large scales ($\gtrsim$100~pc) becomes consistent with a foreground screen (suggesting other geometries tend to be more localized). 

In the scenario in which a foreground-like dust component is dominating the attenuation, the optical depth is expected to behave according to
\begin{equation}
\tau_{n,r}(\lambda) = -\ln \frac{F_n(\lambda)}{F_r(\lambda)} \,,
\end{equation}
where $\tau_{n,r}$ corresponds to the dust optical depth of template $n$, with flux density $F_n(\lambda)$, relative to a reference template $r$, with flux density $F_r(\lambda)$, and it is required that $n>r$ for comparison. From this relation, it is possible to determine the selective attenuation, $Q_{n,r}(\lambda)$, 
\begin{equation}\label{eq:Q_def}
Q_{n,r}(\lambda) = \frac{\tau_{n,r}(\lambda)}{\delta \tau_{Bn,r}^l} \,,
\end{equation}
where $\delta \tau_{Bn,r}^l=\tau_{Bn}^l-\tau_{Br}^l$ is the difference between the Balmer optical depth of template $n$ and $r$. The normalization for the selective attenuation is arbitrary. Following \citet{calzetti94}, we select $Q_{n,r}(5500\mathrm{\AA})=0$ as the zero-point.

In order to construct the templates, we select galaxies in a narrow range of $D_n4000$ values for comparison. This selection removes the systematic trend where galaxies with larger $D_n4000$ values (older ages) are slightly dustier on average, which impacts the inferred attenuation in the UV-optical wavelength range \citep{battisti16}. We select galaxies within a window of $1.13<D_n4000<1.33$, which is centered on the mean value of 1.23 for the full NIR sample. We explored using different selection choices for the sample used in deriving the curve and find that only using $D_n4000$ as a restriction leads to systematically larger inferred attenuation curves in the NIR. The reason for this effect is discussed in \S~\ref{curve_vs_param}. As a demonstration case for now, which will be used for comparison to other sub-populations later, we also impose a restriction on the stellar mass and select the window of $9.1<\log[M_* (M_{\odot})]<9.9$ that is centered on the peak of the distribution. These restrictions limit the sample to 2303 galaxies for the analysis presented in this section.

\begin{figure}
\plotone{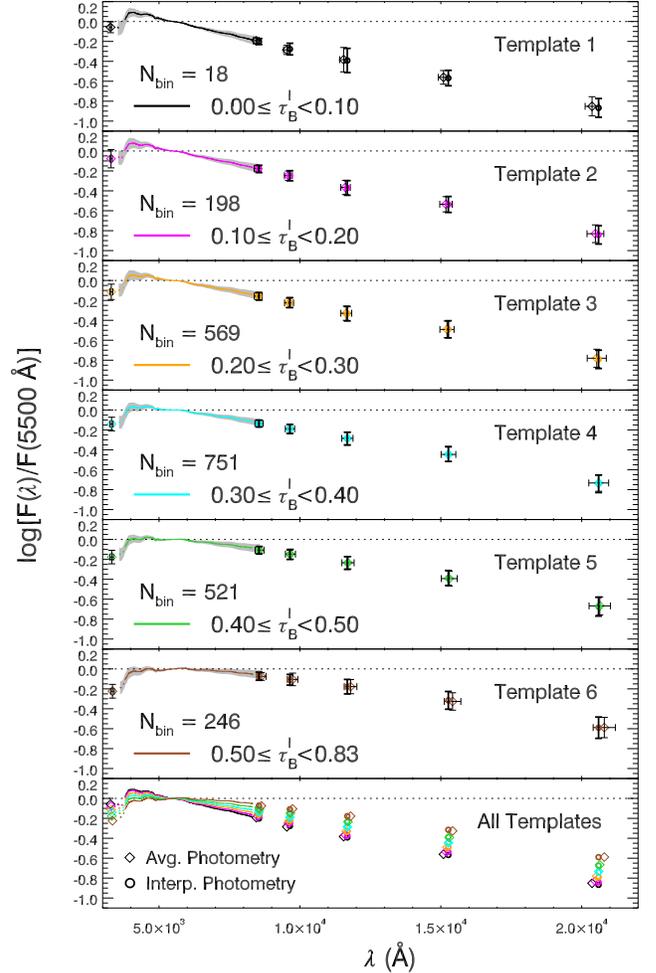}
\caption{Average flux density of galaxies, normalized at 5500~\AA, within each bin of $\tau_B^l$ for the sub-population of galaxies within $1.13<D_n4000<1.33$ \& $9.1<\log[M_* (M_{\odot})]<9.9$ ($N_{\mathrm{tot}}=2303$). The range in $\tau_B^l$ and the number of sources in each bin, $N_{\mathrm{bin}}$, are shown in each panel. The optical measurements are from SDSS spectroscopy. The gray regions denote the area enclosing approximately 68\% of the population. The dotted regions in the optical spectra indicate the average obtained from less than the full sample in that bin (due to varying redshifts), but still containing $>50\%$ of the bin sample. The diamonds show the average flux densities for SDSS $uz$ and NIR $YJHK$ photometry, with errors denoting the 1$\sigma$ range in rest-frame wavelength and flux density values spanned in each bin, respectively. The circles show the average flux densities after interpolating the NIR to fixed effective wavelengths ($\bar{z}=0.06$; see \S~\ref{templates}).  The bottom panel shows a comparison of the average flux density of each bin without the dispersion included for reference. Galaxies with higher $\tau_B^l$ have shallower average SED profiles, which indicates increased levels of dust reddening. \label{fig:F_vs_lam}}
\end{figure}

The galaxies within this sub-population are separated into bins of $\tau_B^l$ and the average SED is determined. We refer the reader to \citet{battisti16} for a description of the technique used for averaging the optical data. For the NIR data, the effective wavelength of the filters vary slightly with the redshift of each galaxy and to correct for this we interpolate the flux density values to a fixed wavelength such that the selective attenuation at each band is determined in a consistent manner. As previously mentioned, the behavior of the NIR spectral region is relatively smooth and this implies that interpolating values for small wavelength offsets is reasonable. The fixed-wavelength flux densities are determined using the observed SDSS $z$, UKIDSS $Y$, $J$, $H$, and $K$ bands and linearly interpolating, in log($F_\lambda$)-log($\lambda$) space, to the effective wavelength corresponding to the average redshift of the sub-population, in this case $\bar{z}=0.06$. For galaxies with $z>\bar{z}$, the $K$ band value is linearly extrapolated (in log-log space) based on the observed $H$, and $K$ band values. The average flux density templates for our galaxies as a function of relative dust attenuation can be seen in Figure~\ref{fig:F_vs_lam}. A clear trend is apparent where galaxies with higher values of $\tau_B^l$ having shallower SED profiles, which is indicative of increasing levels of dust reddening.

A potential concern in this analysis is that $\tau_B^l$ corresponds to the attenuation of ionized gas in star-forming regions and this may be spatially disconnected from the attenuation experienced by the evolved stellar population (which generally dominates the NIR emission). With our current data, there is no direct way to break the differential attenuation of old and young stellar populations and this will introduce systematic uncertainties. This issue is is likely to be suppressed to some extent by our sample selection, which primarily corresponds to actively star-forming galaxies with young average stellar population ages. For example, galaxies with $D_n4000\sim1.3$ correspond to a mean stellar population age of $1-3$~Gyr \citep{kauffmann03a}. Therefore, it can be expected that some fraction of the NIR SED in our galaxies will still be associated with current star formation and would therefore be affected by dust in manner correlated with $\tau_B^l$. The existence of the trend in Figure~\ref{fig:F_vs_lam} extending out to our longest wavelength band suggests that $\tau_B^l$ does appear to quantify the attenuation for NIR light in our sample. However, we acknowledge this as a source of uncertainty in our analysis.

\begin{figure*}$
\begin{array}{rc}
\includegraphics[width=3.4in]{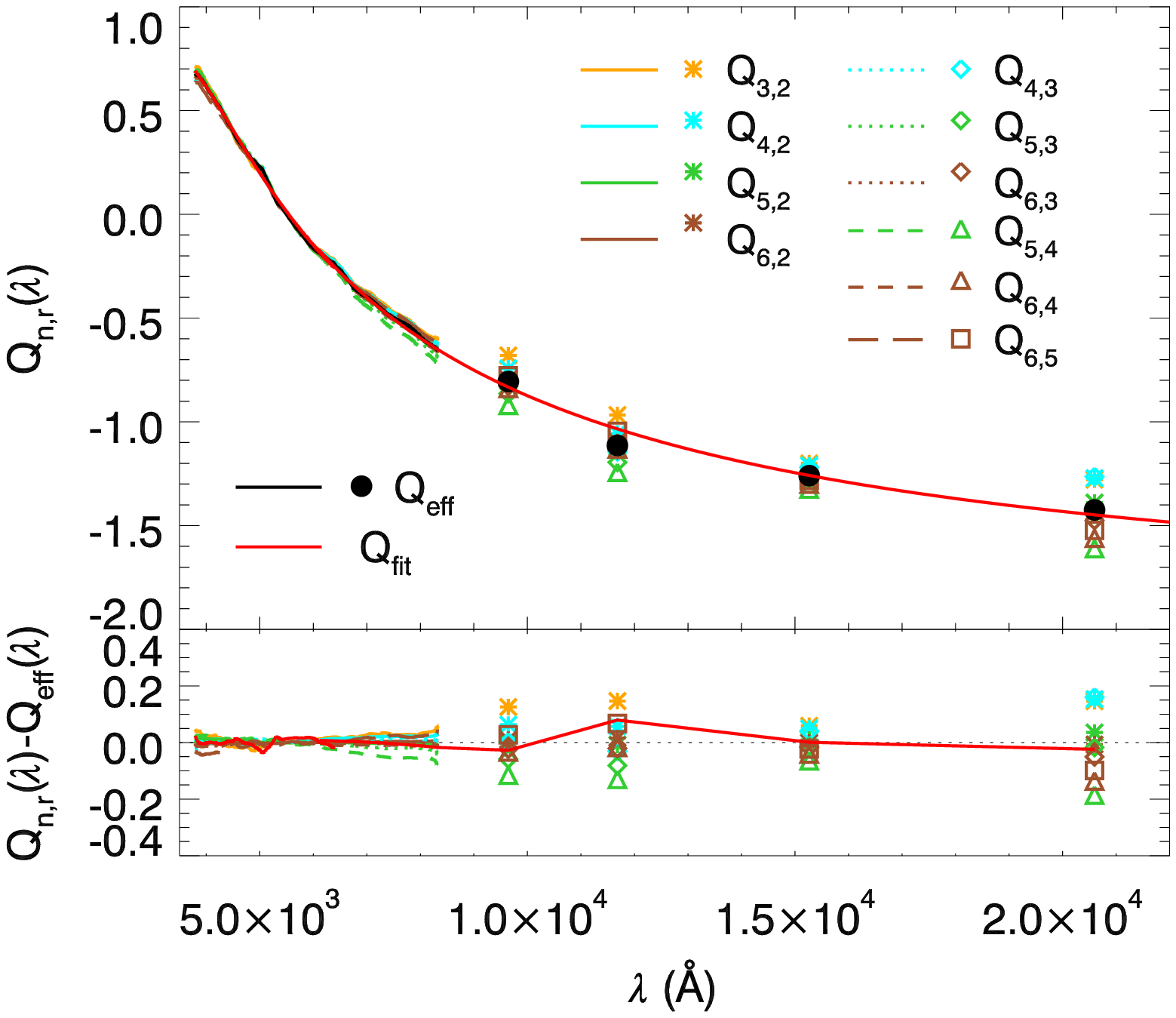} &
\includegraphics[width=3.4in]{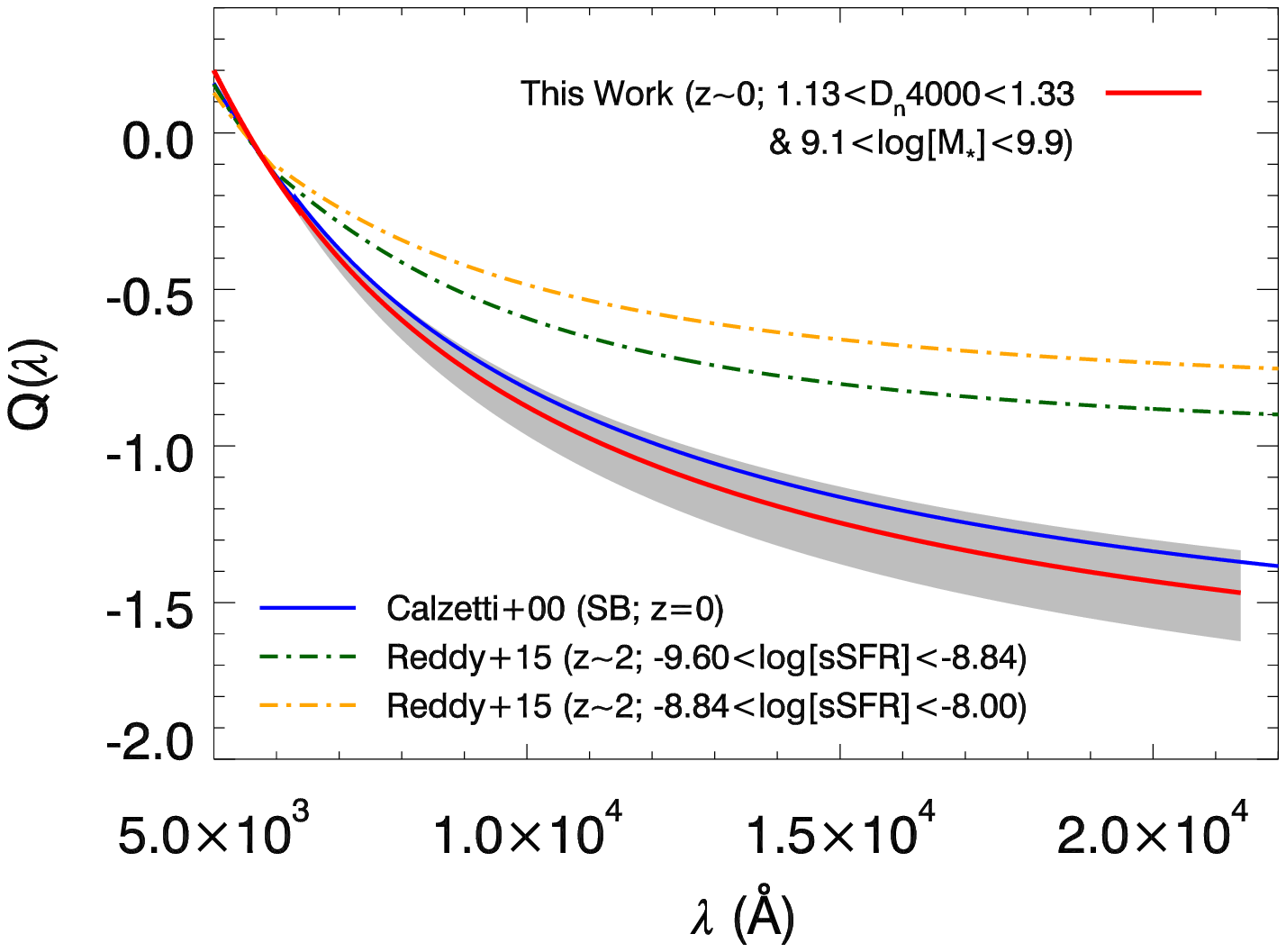} \\
\end{array}$
\caption{\textit{Left:} Selective attenuation curve, $Q_{n,r}(\lambda)$, for a subset of our sample based on comparing a given template, $n$, to a reference template, $r$, at lower $\tau_B^l$. Also shown is the effective curve, $Q_{\mathrm{eff}}(\lambda)$ (solid black line and circles), which is the average value of $Q_{n,r}(\lambda)$. The solid red line consists of two polynomial fits to $Q_{\mathrm{eff}}(\lambda)$ determined by separately fitting the optical ($3800-8325$~\AA) and NIR ($6000-22000$~\AA) regions. The lower panel shows the difference between each curve relative to $Q_{\mathrm{eff}}(\lambda)$. \textit{Right:} Comparison between our selective attenuation curve and those in the literature. The gray region denotes the range of $Q_{n,r}(\lambda)$ values. The solid blue line is the starburst selective attenuation curve of \citet{calzetti00} and the dash-dot lines are the curves of $z\sim2$ SFGs from \citet{reddy15} divided according to sSFR. The selective attenuation curve for our sub-population lies below the starburst relation in the NIR.
\label{fig:Q_eff}}
\end{figure*}

We show the selective attenuation curves that are obtained using various template combinations in Figure~\ref{fig:Q_eff}. We do not use template 1 in the analysis for this sub-population because it has a relatively small sample size and this makes it more susceptible to variation in the stellar population age. The effective attenuation curve, $Q_{\mathrm{eff}}(\lambda)$, for the sample is found by taking the average value of $Q_{n,r}(\lambda)$ from templates 2-6. The NIR portion of $Q_{\mathrm{eff}}(\lambda)$ is fit using the SDSS data longward of 6000~\AA\ combined with the NIR bands. We use the SDSS spectroscopic range ($3800-8325$~\AA) to fit the optical region of the NIR sub-population, which is used later to normalize the curve for differential reddening. Fitting the NIR portion of $Q_{\mathrm{eff}}(\lambda)$ to a single second-order polynomial as a function of $x=1/\lambda$~($\micron^{-1}$) results in
\begin{multline}\label{eq:Q_fit}
Q_{\mathrm{fit}}(x) = -1.996 + 1.135x - 0.0124x^2\,,\\ 0.63~\micron\le\lambda<2.1~\micron \,. 
\end{multline}

We compare our updated selective attenuation curve to others in the literature in Figure~\ref{fig:Q_eff}. To give a sense of the uncertainty, a gray region denoting the range of $Q_{n,r}(\lambda)$ from the various templates is shown (i.e., region spanned by the data shown in Figure~\ref{fig:Q_eff}). We include the curves of local starburst galaxies from \citet{calzetti00} and higher redshift ($z\sim2$) SFGs from \citet{reddy15}. The selective attenuation curves of \citet{reddy15} are divided according to sSFR. The selective attenuation curve for our sub-population lies below the local starburst relation \citep{calzetti00} in the NIR.

The selective attenuation is related to the total-to-selective attenuation, $k(\lambda)$, through the following relation
\begin{equation}\label{eq:k_def}
k(\lambda)= fQ(\lambda)+R_V \,,
\end{equation}
where $f$ is a constant required to make $k(B)-k(V)\equiv1$,
\begin{equation}\label{eq:f_def}
f= \frac{1}{Q_{\mathrm{eff}}(B)-Q_{\mathrm{eff}}(V)} \,,
\end{equation}
and $R_V$ is the total-to-selective attenuation in the $V$ band, which is the vertical offset of the curve from zero at 5500~\AA. We assume $B$ and $V$ bands to be 4400~\AA\ and 5500~\AA, respectively. The term $f$ is accounting for differences in the reddening between the ionized gas and the stellar continuum and can also be expressed as 
\begin{equation}
f = \frac{k(H\beta)-k(H\alpha)}{E(B-V)_{\mathrm{star}}/E(B-V)_{\mathrm{gas}}} \,,
\end{equation}
where $k(H\beta)$ and $k(H\alpha)$ are the values for the intrinsic extinction curve of the galaxy and \textit{not} from the attenuation curve. As such, the quantity $fQ(\lambda)$ represents the true wavelength-dependent behavior of the attenuation curve on the stellar continuum.

\begin{figure}
\plotone{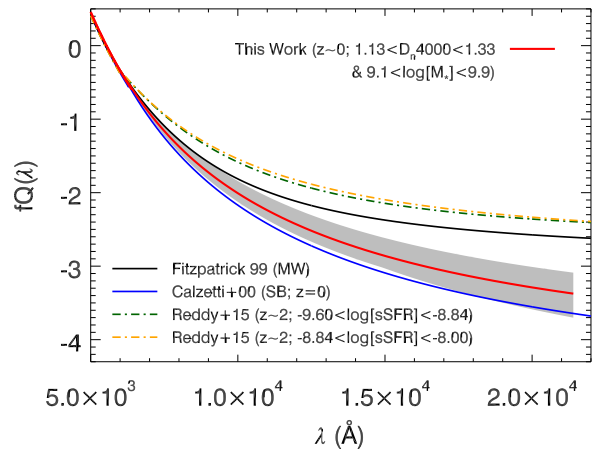}
\caption{Normalized selective attenuation curve $fQ(\lambda)$ derived from a subset of our sample compared to the literature. The term $f$ is required to make the curve have $k(B)-k(V)\equiv1$. Lines are the same as in Figure~\ref{fig:Q_eff}, but with the addition of the MW extinction curve \citep{fitzpatrick99} (solid black). The gray region denotes the range of $fQ_{n,r}(\lambda)$ values, where $f$ varies in each case. After applying $f$, the curve is above the local starburst relation \citep{calzetti00}. \label{fig:fQ_compare}}
\end{figure}

The value of $f$ for the average selective attenuation curve for this sub-population, based on the fit to the optical region, $Q_{\mathrm{fit}}(\lambda)$, is 2.30$\substack{+0.09 \\ -0.10}$, where the uncertainty here reflects the maximum and minimum values from fits using individual $Q_{n,r}(\lambda)$. We compare our selective attenuation curve with this factor included to the curves in the literature in Figure~\ref{fig:fQ_compare}. For reference, the values of $f$ are $f=2.66$ in \citet{calzetti00}, and $f=2.68$ and 3.18 for the lower and higher sSFR sub-population, respectively, in \citet{reddy15}. After applying $f$, our curve is between the local starburst relation \citep{calzetti00} and the MW extinction curve.

\subsection{Curve Normalization, \texorpdfstring{$R_V$}{RV}}\label{normalization}
Determination of the normalization, $R_V$, can be performed in two distinct ways: (1) measuring NIR photometry or spectroscopy to wavelengths where the total-to-selective attenuation approaches zero, $k(\lambda\rightarrow\infty)\sim0$, which occurs at $\lambda\sim2.85$~\micron\ \citep[e.g., ][]{calzetti97b,gordon03,reddy15}, although we note that studies of the MW extinction curve suggest that the zero-point occurs at longer wavelengths \citep[e.g.,][and references therein]{xue16}, or (2) measuring the total infrared luminosity to infer the total dust attenuation under the assumption of energy balance \citep[e.g.,][]{calzetti00}.

\begin{figure*}$
\begin{array}{cc}
\includegraphics[width=3.3in]{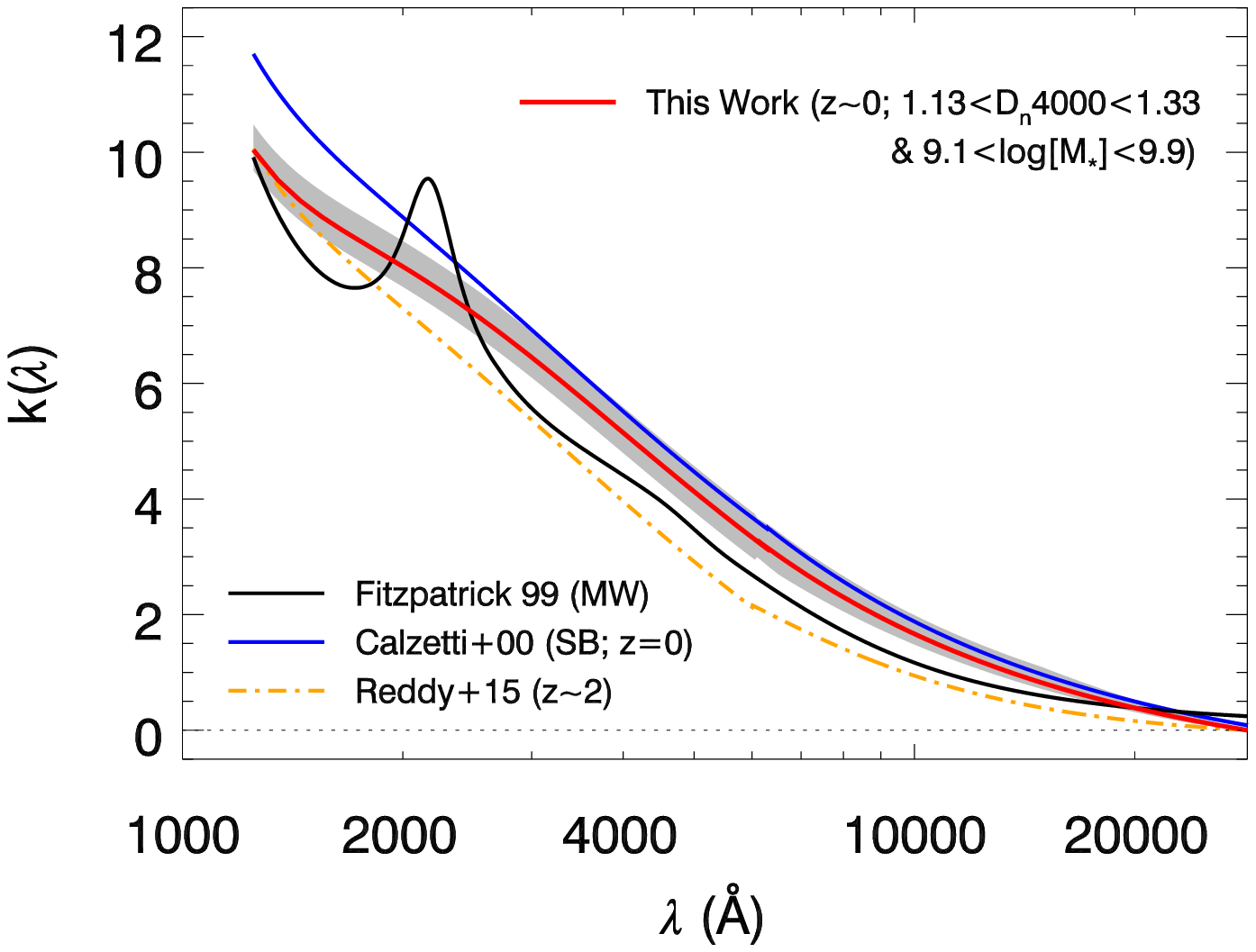} &
\includegraphics[width=3.3in]{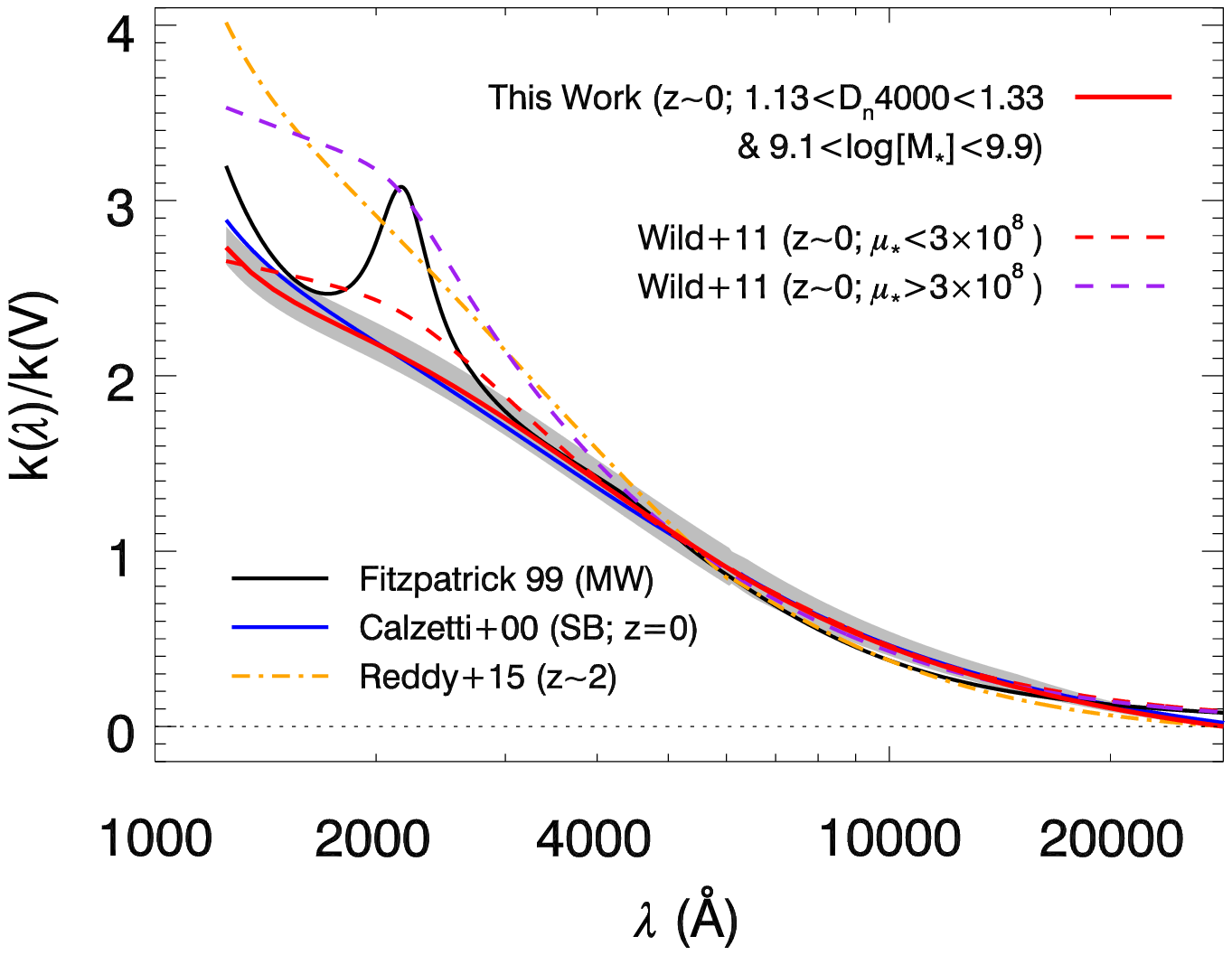} \\
\end{array}$
\caption{\textit{Left:} The total-to-selective dust attenuation, $k(\lambda)$, derived from a subset of our sample compared to the literature. The curve for SFGs has a lower total attenuation in the FUV than local starburst galaxies by about 15\%, but becomes similar at longer wavelengths. The gray region demonstrates the uncertainty associated with the $R_V$ value. \textit{Right:} Similar to the previous plot, but with the curve normalized at $k(V)$, which allows comparison with the curves of \citet{wild11} for local SDSS galaxies (divided according to stellar mass surface density, $\mu_*$; dashed lines).
\label{fig:k_compare}}
\end{figure*}

We first explore using method (1) to determine $R_V$. It can be seen in Figure~\ref{fig:fQ_compare} that the range of $fQ_{n,r}(\lambda)$ values (gray region) in the NIR is significant after multiplying by $f$, unlike what occurred at UV-optical wavelengths \citep{battisti16}, and this indicates that the estimate of $R_V$ will have significant uncertainty. Part of this uncertainty is an extension of the uncertainty in $f$, which has a stronger impact on the curve the further away one goes from the normalization point ($B-V$), but it would also reflect variation in this value across the sample (see \S~\ref{curve_vs_param}). Extrapolating for the value of $fQ_{\mathrm{fit}}(\lambda)$ out to 2.85$~\micron$ using Equation~(\ref{eq:Q_fit}) and forcing $k(2.85~\micron)=0$ gives a value of $R_V=3.67\substack{+0.44 \\ -0.35}$. Using this normalization, we can define the total attenuation curve for local SFGs as 

\begin{equation}\label{eq:k_lambda}
  \begin{array}{ccr}\vspace{2pt}
k(\lambda)& \multicolumn{2}{c}{= 2.40(-2.488 + 1.803x - 0.261x^2 + 0.0145x^3)}\\ \vspace{5pt}
 & \hspace{20pt} +3.67 & 0.125~\micron\le\lambda<0.63~\micron \, \\\vspace{2pt}
 & \multicolumn{2}{c}{= 2.30(-1.996 + 1.135x - 0.0124x^2) +3.67 }\\ 
 &  & 0.63~\micron\le\lambda<2.1~\micron \,. 
\end{array}
\end{equation}
We will refer to the curve given by Equation~(\ref{eq:k_lambda}) as our fiducial case throughout the rest of the paper. If, instead of 2.85$~\micron$, we extrapolate our curve to infinite wavelength and force $k(\lambda\rightarrow\infty)=0$, we find $R_V=4.58\substack{+0.87 \\ -0.69}$. Given the larger uncertainties associated with extrapolating to infinite wavelength, we are inclined to support the first method, as it does not rely on extrapolating too far from where we have constraining information ($\lambda\sim2.1~\micron$).

The attenuation curve given by Equation~(\ref{eq:k_lambda}) is shown in comparison to others curves in Figure~\ref{fig:k_compare}. In this figure we also compare to the results of \citet{wild11} based on local SDSS galaxies, which were divided according to stellar mass surface density, $\mu_*$, with the break corresponding to the value which separates the bimodal local galaxy population into bulge-less ($\mu_*<3\times10^8~M_\odot~\mathrm{kpc}^{-2}$) and bulged ($\mu_*>3\times10^8~M_\odot~\mathrm{kpc}^{-2}$) galaxies \citep{kauffmann03b}. The curves of \citet{wild11} were further sub-divided according to sSFR and axial ratio ($b/a$). For clarity, we only reproduce the curves corresponding to the average sample value ($\log[\mathrm{sSFR~(yr}^{-1})]=-9.5$ and $b/a=0.6$) in that work. We note that the curves from \citet{wild11} are presented in a manner that makes direct comparison in previous figures unclear and they were therefore not shown. We find that our curve is similar to the lower $\mu_*$ sample of \citet{wild11}, which is expected given the overlap in these datasets (our sample is predominantly spiral galaxies that would be classified into the lower $\mu_*$ sample). We refer the reader to \citet{battisti16} for a discussion of the differences in the UV shape.

As a check on the normalization found above, we would nominally use method (2) as an independent test. Unfortunately, this sample lacks far-IR data necessary for accurate estimates of the total infrared luminosities. Regardless, it is still possible to check if the normalization found from method (1) is consistent with other samples of local SFGs for which the necessary data is available for an energy balance comparison. We perform this analysis in \S~\ref{energy_balance}.

\subsection{The Variation of the Attenuation Curve}\label{curve_vs_param}
In this section, we examine the differences in the attenuation curve and its normalization as a function of galaxy properties. Such variations may arise as a result of differences in the size distributions of dust grains \citep[e.g.,][]{cardelli89,weingartner&draine01} or the average geometry between the stars and dust \citep[e.g.,][]{calzetti01} as a function of certain galactic properties. For example, the value of $R_V$ in the MW extinction curve shows strong evolution with the density of the interstellar medium, such that denser environments have larger values of $R_V$ due to an increase in the fraction of large dust grains \citep[e.g.,][]{weingartner&draine01}. However, it should be noted that $R_V$ in an attenuation curve is fundamentally different from an extinction curve because of the additional geometric and scattering components in the case of attenuation \citep[see also][]{calzetti00}. We refer the reader to \citet{battisti16} for additional discussion on the influence of galaxy properties on dust attenuation.

\begin{figure*}
\begin{center}
\includegraphics[width=6.8in]{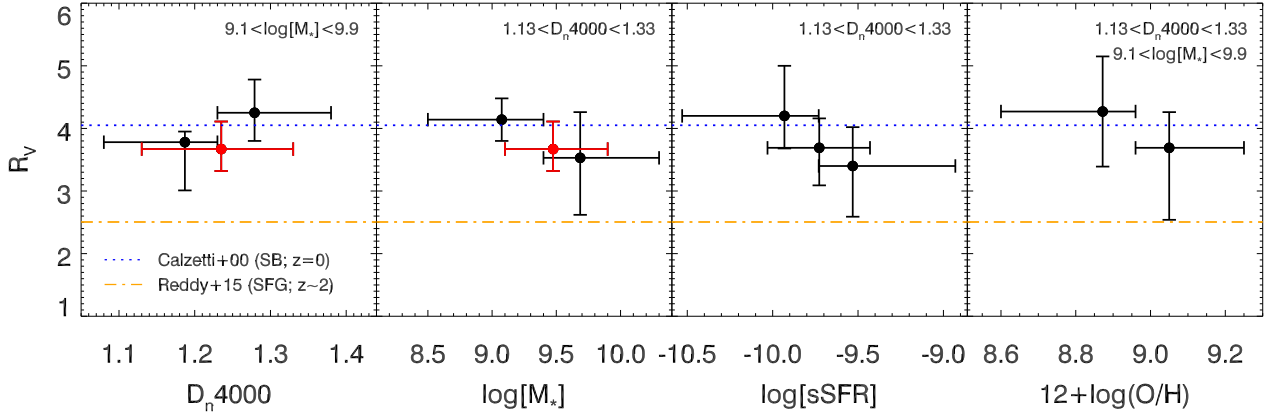}
\end{center}
\caption{Comparison of the normalization factor, $R_V$, determined from the selective attenuation curve of sub-populations of galaxies divided according to $D_n4000$, sSFR, stellar mass, and gas-phase metallicity (12+log(O/H)) compared to the fiducial case ($1.13<D_n4000<1.33$ \& $9.1<\log[M_* (M_{\odot})]<9.9$; red symbol). The x-axis value corresponds to the median in the sub-population, with the errorbar reflecting the range spanned. The y-axis uncertainties correspond to the range of $R_V$ from extrapolating $k(2.85~\micron)=0$. The sub-populations show tentative trends but generally appear consistent given the uncertainties.  
 \label{fig:Rv_compare_subsample}}
\end{figure*}

\begin{table*}
\begin{center}
\caption{Fit Parameters of $Q(\lambda)$ in the NIR as a Function of Galaxy Properties \label{Tab:Q_vs_param}}
\begin{tabular}{ccccccc}
\hline\hline 
      $x$    &     range   & $f$ &  $p_0$    &  $p_1$    & $p_2$ &  $R_V(2.85~\micron)$  \\  \hline 

$D_n4000$ & $1.08<x<1.23^a$ & 2.34$\substack{+0.80 \\ -0.32}$ & -1.971 & 1.005 & 5.640$\times$10$^{-2}$ & 3.78$\substack{+0.17 \\ -0.77}$ \\ 
          & $1.23<x<1.38^a$ & 2.67$\substack{+1.25 \\ -0.59}$ & -2.070 & 1.422 & -1.513$\times$10$^{-1}$ & 4.25$\substack{+0.53 \\ -0.45}$ \\ \\

log[sSFR (yr$^{-1}$)] & $-10.53<x<-9.73^b$ & 1.94$\substack{+0.36 \\ -0.18}$ & -2.820 & 1.958 & -2.144$\times$10$^{-1}$ & 4.20$\substack{+0.80 \\ -0.52}$  \\
          & $-10.03<x<-9.43^b$ & 1.81$\substack{+0.46 \\ -0.16}$ & -2.594 & 1.668 & -1.291$\times$10$^{-1}$ & 3.69$\substack{+0.47 \\ -0.60}$ \\
          & $-9.73<x<-8.93^b$ & 1.69$\substack{+0.46 \\ -0.20}$ & -2.537 & 1.515 & -6.611$\times$10$^{-2}$ & 3.40$\substack{+0.62 \\ -0.81}$ \\ \\

log[$M_* (M_{\odot})$] & $8.5<x<9.4^b$ & 2.36$\substack{+1.16 \\ -0.60}$ & -2.238 & 1.394 & -7.569$\times$10$^{-2}$ & 4.14$\substack{+0.34 \\ -0.34}$  \\
           & $9.1<x<9.9^b$  & 2.30$\substack{+0.09 \\ -0.10}$ & -1.996 & 1.135 & -1.238$\times$10$^{-2}$ & 3.67$\substack{+0.44 \\ -0.35}$ \\
          & $9.4<x<10.3^b$ & 2.53$\substack{+0.41 \\ -0.32}$ & -1.726 & 9.435$\times$10$^{-1}$ & 6.201$\times$10$^{-3}$ & 3.53$\substack{+0.73 \\ -0.91}$ \\ \\

12+log(O/H) & $8.60<x<8.96^{ab}$ & 2.22$\substack{+1.00 \\ -0.58}$ & -2.483 & 1.640 & -1.439$\times$10$^{-1}$ & 4.27$\substack{+0.88 \\ -0.88}$ \\
          & $8.96<x<9.25^{ab}$ & 2.76$\substack{+1.70 \\ -0.41}$ & -1.646 & 8.697$\times$10$^{-1}$ & 2.438$\times$10$^{-2}$ & 3.69$\substack{+0.57 \\ -1.15}$  \\ \hline 
\end{tabular}
\end{center}
\textbf{Notes.} The uncertainty in $f$ denotes the maximum and minimum values from fits using individual $Q_{n,r}(\lambda)$ for each sub-population (see \S~\ref{templates}). The functional form of these fits are $Q(\lambda) = p_0+p_1x+p_2x^2$, where $x=1/\lambda$~($\micron^{-1}$). The normalization $R_V(2.85~\micron)$ is determined by extrapolating this function out to 2.85~$\micron$ and forcing $k(2.85~\micron)=0$ (see \S~\ref{normalization}). $^a$This also has the constraint that $9.1<\log[M_* (M_{\odot})]<9.9$. $^b$This also has the constraint that $1.13<D_n4000<1.33$.
\end{table*}

We examine the behavior of the attenuation curve when dividing the sample according to $D_n4000$, sSFR, stellar mass, and metallicity. The galaxy inclination has also been observed to influence the attenuation of galaxies \citep[e.g.,][]{wild11} and is expected from theory \citep[e.g.,][]{tuffs04,pierini04,chevallard13,natale15}, but we defer exploring this parameter for our sample to a subsequent paper in order to perform a more in-depth analysis. We follow the same procedure outlined in \S~\ref{templates} and divide each sample into 6 bins of $\tau_B^l$ and construct average flux templates. Similar to before, we do not utilize bins with less than 100 galaxies for determining the attenuation curve, but this only acts to exclude a single bin in any given sub-population. The mean redshift of each sub-population is used to set the effective wavelengths for the NIR interpolation (usually $\bar{z}\sim0.06$). It is important to stress that this analysis does not provide information on the extent of variation at the individual galaxy level and only provides insight into systematic differences among the galaxy populations taken in average.

Our approach is to separate the sample according to various properties while also using the constraint that $1.13< D_n4000 <1.33$, in order to limit stellar population age effects. The effective attenuation curve, $Q_{\mathrm{eff}}(\lambda)$, in all cases is well approximated by a single second-order polynomial. The values of these fits are presented in Table~\ref{Tab:Q_vs_param}.  The sub-populations show differences in $f$ (or equivalently in $E(B-V)_{\mathrm{star}}/E(B-V)_{\mathrm{gas}}$), which may result from changes in the relative contribution to the global flux density from massive stars between cases. The normalization for each sub-population is shown in Figure~\ref{fig:Rv_compare_subsample}.

First, we examine the result of taking sub-populations within $D_n4000$. We find that separating by $D_n4000$ alone (not shown) results in a systematically higher inferred attenuation in the NIR than is found when also adding a constraint in stellar mass, $9.1<\log[M_* (M_{\odot})]<9.9$ (or in sSFR). Given that the strength of $D_n4000$ is dominated by intermediate mass stars (low mass-to-light ratio component), we expect that it does a relatively poor job at discriminating stellar population ages older than $\sim$1~Gyr, which has an increasingly important impact at longer wavelengths in the NIR where the stellar continuum can be dominated by low mass stars (high mass-to-light ratio component). For this reason, we believe that using $D_n4000$ alone creates a bias in our derived attenuation curve in the NIR. Additionally separating the sample by stellar mass or sSFR should provide a better way of selecting galaxies with a similar contribution from older ($>$1~Gyr) stars. The sub-populations with this additional restriction appear to be consistent with the fiducial case (red symbol).

Next, we examine the results of separating the sample by sSFR and stellar mass. For sSFR, we also checked the result of taking a window in sSFR centered on the peak of the distribution and  found that it gives a very similar result to the what is found from the fiducial case. There are weak indications that higher sSFR (or lower stellar mass) cases may favor a lower curve normalization ($R_V$), however given their uncertainties this is not very significant and they can be viewed as consistent with the range found for the fiducial case. 

Last, we examine separating the sample by metallicity. In order to limit age effects, we impose restrictions on both stellar mass and $D_n4000$. It is important to note that galaxies in the SDSS spectroscopic sample with metallicity measurements are primarily located at $8.6<12+\log(O/H)<9.25$, implying that we cannot constrain the behavior of metal-poor galaxies. We find that these sub-populations have particularly high uncertainties, reflecting smaller number statistics, but are also in agreement with the fiducial case. 

From these results, it appears that the normalization $R_V$ does not vary in a significant manner as a function of $D_n4000$, sSFR, stellar mass, and metallicity over the ranges probed by our sample. We stress that the potential for systematic effects on the inferred attenuation curve increases when using these smaller sample sizes, which leads us to further caution an over-interpretation of the observed trends with properties. Expanding on these studies to higher redshifts to explore a wider range of parameter space in galaxy properties will help to determine if these effects are real. For now, we suggest that that our fiducial curve is reasonable in describing the attenuation in a majority of our sample.

\section{Energy Balance Comparison}\label{energy_balance}
The IR emission of SFGs, assuming minimal AGN influence, arises primarily from the thermal radiation of dust grains heated by the absorption of UV-to-NIR photons from stars. In this scenario, one expects that the amount of energy absorbed in the UV-to-NIR by dust should be balanced by its IR emission due to energy conservation. There are two components to consider for the absorption of UV-to-NIR photons; the stellar continuum from $0.0912<\lambda<2.85~\micron$ which will be characterized by the attenuation curve, and ionizing photons ($\lambda<0.0912~\micron$) absorbed by dust either directly or after they have been reemitted by hydrogen through recombination lines. In practice, this comparison is quite difficult owing to the large uncertainty in calculating the fraction of ionizing photons directly absorbed by dust \citep[e.g.,][]{inoue01a,hirashita03}.

To perform this analysis we need a dataset with photometric data from the UV-to-far-IR, as well as optical spectroscopy to determine the attenuation from the Balmer decrement ($\tau_B^l$). We utilize the Galaxy Multiwavelength Atlas from Combined Surveys \citep[GMACS;][]{johnson07a,johnson07b} dataset, which has a publicly available catalog\footnote{\url{http://user.astro.columbia.edu/~bjohnson/GMACS/catalogs.html}} of cross-matched photometric measurements from \textit{GALEX}, SDSS (with spectroscopy), 2MASS, and \textit{Spitzer}. The GMACS sample consist of the Lockman Hole, the Spitzer First Look Survey, and the SWIRE ELAIS-N1 and N2 fields. The photometry for all bands has been performed to represent the global values for each galaxy \citep[see][for details]{johnson07a}. The optical emission lines measurements and galaxy properties are obtained from the DR7 of the MPA/JHU group. We select galaxies identified as SFGs using the same emission line diagnostics as our parent sample using lines with $S/N>5$. In addition, we require $\tau_B^l>0$ and full MIPS coverage (24, 70, $160~\micron$) to ensure an accurate measurement of \ltir\ (the dust emission peak typically occurs between 70-$160~\micron$). This leaves a final sample of 166 galaxies for analysis. We also examined the possibility of using various large-area \textit{Herschel} surveys, but found that very few of these surveys have deep observations with PACS which are necessary to get accurate \ltir\ estimates at these low redshifts \citep{galametz13} and they were therefore excluded.

The amount of stellar continuum light that is absorbed by dust in the wavelength range of $0.0912<\lambda<2.85~\micron$ is determined using the difference between the total intrinsic and observed luminosity, 
\begin{equation}
\Delta L_{\mathrm{UV-NIR}} = L_{\mathrm{int}}(0.0912-2.85)-L_{\mathrm{obs}}(0.0912-2.85) \,. 
\end{equation}
The observed SED, $F_{\mathrm{obs}}(\lambda)$, is determined by fitting the global photometry using the SED-fitting code EAZY \citep{brammer08} with the redshift fixed to its spectroscopically determined value. The intrinsic SED, $F_{\mathrm{int}}(\lambda)$, is determined by correcting $F_{\mathrm{obs}}(\lambda)$ using the attenuation curve given by Equation~(\ref{eq:k_lambda}), 
\begin{equation}
F_{\mathrm{int}}(\lambda) = F_{\mathrm{obs}}(\lambda) 10^{0.4E(B-V)_{\mathrm{star}}k(\lambda)} \,,
\end{equation}
where the color excess of the stellar continuum can be determined from the Balmer optical depth (combining Equation~\ref{eq:EBV_gas} \& \ref{eq:f_def})
\begin{equation}\label{eq:EBV_star}
E(B-V)_{\mathrm{star}} = \frac{1.086 \tau_B^l}{f} \,,
\end{equation}
and we use $f=2.40$ corresponding to the average value for SFGs from \citet{battisti16}. We will explore the result of changing the assumed value of $R_V$ and $f$ below. For this analysis we are extrapolating the UV-optical part of Equation~(\ref{eq:k_def}) down to $0.0912~\micron$. Recent results by \citet{reddy16} suggest that the actual shape of the attenuation curve may be shallower than this extrapolation. We have examined the impact of using a shallower slope, similar in shape to their updated \citet{calzetti00} value, and find that it typically changes the value of $\Delta L_{\mathrm{UV-NIR}}$ by less than a few percent. We convert these SEDs to an integrated luminosity using the luminosity distance, $L_{\mathrm{rest}}(0.0912-2.85)=4\pi D_{\mathrm{lum}}^2\int(\nu F_\nu(\lambda))_{\mathrm{obs}}d\nu$.

The fraction of ionizing photons directly absorbed by dust, ($1-f_{\mathrm{ion}}$), cannot be constrained for galaxies in general. Measurements of the fraction of emerging flux, $f_{\mathrm{ion}}$, for individual HII regions indicate that it can vary substantially, with a range of $0.2\lesssim f_{\mathrm{ion}}\lesssim1.0$ within local galaxies \citep{inoue01a,inoue01b}. However, galaxy-wide averages for local spiral galaxies (MW, M31, M33) suggest a typical value of $f_{\mathrm{ion}}\sim0.4$ \citep{inoue01b}. Given that the galaxies selected for this analysis correspond to actively star-forming systems (which are preferentially disk galaxies), we adopt $f_{\mathrm{ion}}=0.4$ as a reasonable choice moving forward. However, we revisit this assumption later on. The intrinsic luminosity of ionizing photons, $L_{\mathrm{ion}}$, is 
\begin{equation}
L_{\mathrm{ion}} = \frac{h\langle\nu_{\mathrm{ion}}\rangle Q(\mathrm{H}^0)}{f_{\mathrm{ion}}}  \,,
\end{equation}
where $h\langle\nu_{\mathrm{ion}}\rangle$ is the average energy of an ionizing photon, $Q(\mathrm{H}^0)$ is the apparent ionizing photon rate, and dividing by $f_{\mathrm{ion}}$ converts the apparent rate into the value expected in the absence of direct dust absorption. We assume an average photon energy of $3.06\times10^{-11}$~erg ($\langle\lambda_{\mathrm{ion}}\rangle=650$~\AA), which corresponds to the effective value obtained using Starburst99 \citep{leitherer99} stellar population models with continuous star formation over timescales longer than 10~Myr \citep[we assume solar metallicity and the IMF of][]{kroupa01}. Although we find that adopting higher or lower average photon energies only effects the derived energy balance ratios at the level of a few percent (due to a majority of the energy absorbed by dust occurring at non-ionizing wavelengths). The ionizing photon rate is determined from the observed H$\alpha$ luminosity assuming case-B recombination with $T_{\mathrm{e}}=10^4$~K and $n_{\mathrm{e}}=100$~cm$^{-3}$ \citep{osterbrock06}
\begin{equation}
Q(\mathrm{H}^0) = \frac{L_{\mathrm{H}\alpha,\mathrm{obs}}\times 10^{0.4E(B-V)_{\mathrm{gas}}k(H\alpha)}}{1.37\times10^{-12}}~ \mathrm{s}^{-1} \,,
\end{equation}
where we are correcting for dust extinction using the measured Balmer decrement (Equation~\ref{eq:EBV_gas}) with a MW curve, and the luminosity is in units of erg~s$^{-1}$. We have applied an aperture correction to obtain total estimates from the fiber values using the prescription from \citet{hopkins03}, which uses the difference between the $r$-band Petrosian and fiber magnitudes. From these we finally obtain the total luminosity of ionizing photons directly absorbed by dust, $\Delta L_{\mathrm{ion}}$,
\begin{equation}
\Delta L_{\mathrm{ion}} = (1-f_{\mathrm{ion}}) L_{\mathrm{ion}} \,.
\end{equation}
For reference, we find that the ratio of $L_{\mathrm{ion}}  / L_{\mathrm{UV-NIR}}$ and $\Delta L_{\mathrm{ion}}  / \Delta L_{\mathrm{UV-NIR}}$ have median values of 14\% and 24\%, respectively. We neglect the contribution to the total attenuated flux of the dust-absorbed portion of the lines in the hydrogen cascade, since this represents less than a few percent of $\Delta L_{\mathrm{ion}}$. 

The total infrared luminosity, \ltir, for each galaxy is calculated using the prescription from \citet{draine&li07}, which uses IRAC and MIPS bands (8, 24, 70, $160~\micron$; their Equation~22) from \textit{Spitzer}. This formulation was constructed to match their model-based dust SEDs. The infrared luminosity range for GMACS galaxies using this method is $2.7\times10^{41}<L_{\mathrm{TIR}}/(\mathrm{erg\ s}^{-1})<4.5\times10^{44}$.

\begin{figure}
\begin{center}
\includegraphics[width=3.5in]{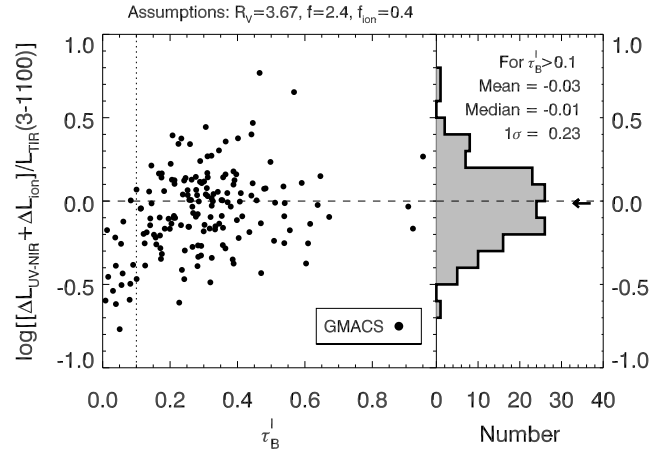}
\end{center}
\caption{Energy balance comparison for 166 SFGs in the GMACS sample. Assuming energy conservation, the loss of energy in the UV-to-NIR will be balanced by the emission in the IR ($[\Delta L_{\mathrm{UV-NIR}} + \Delta L_{\mathrm{ion}}]/L_{\mathrm{TIR}}=1$). The main assumptions made for this analysis are shown at the top of the figure and described in \S~\ref{energy_balance}. The median value for the distribution (black arrow) is close to zero, in log-space, indicating general agreement but with significant scatter ($1\sigma=0.23$). This large variation may correspond to differences in the dust attenuation properties of SFGs. \label{fig:energy_ratio_compare}}
\end{figure}

\begin{figure*}
\begin{center}
\includegraphics[width=7in]{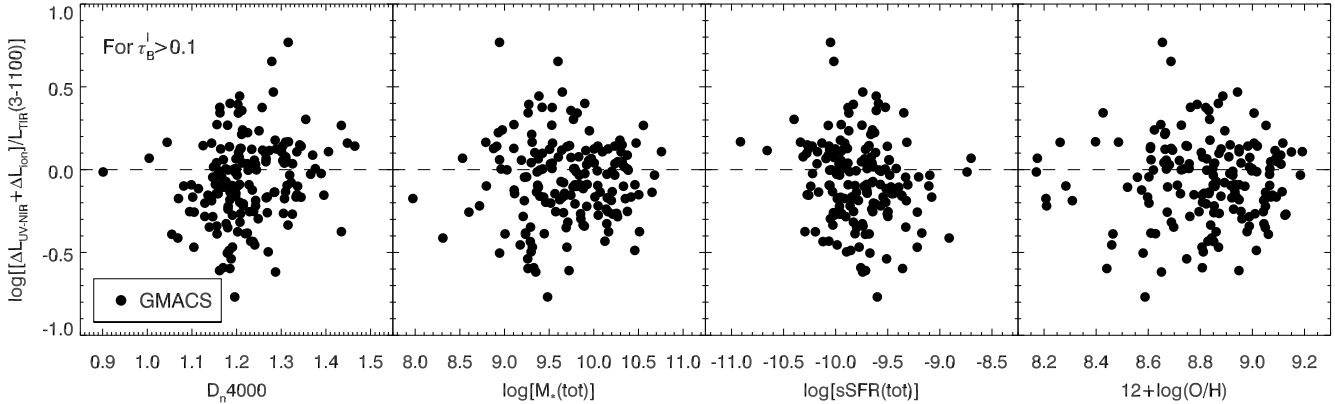}
\end{center}
\caption{Energy balance comparison for SFGs in the GMACS sample as a function of the galaxy parameter. We only consider the galaxies with $\tau_B^l>0.1$, where the attenuation probed by $\tau_B^l$ should provide a significant portion of the observed $L_{\mathrm{TIR}}$. We use the total stellar mass and sSFR, as opposed to a fiber value, here because the energy balance analysis is based on the total flux measurements for these galaxies. No obvious trends are observed among the parameters.
 \label{fig:GMACS_param_vs_energy_ratio}}
\end{figure*}

\begin{figure*}
\begin{center}$
\begin{array}{lcr}
\includegraphics[width=2.2in]{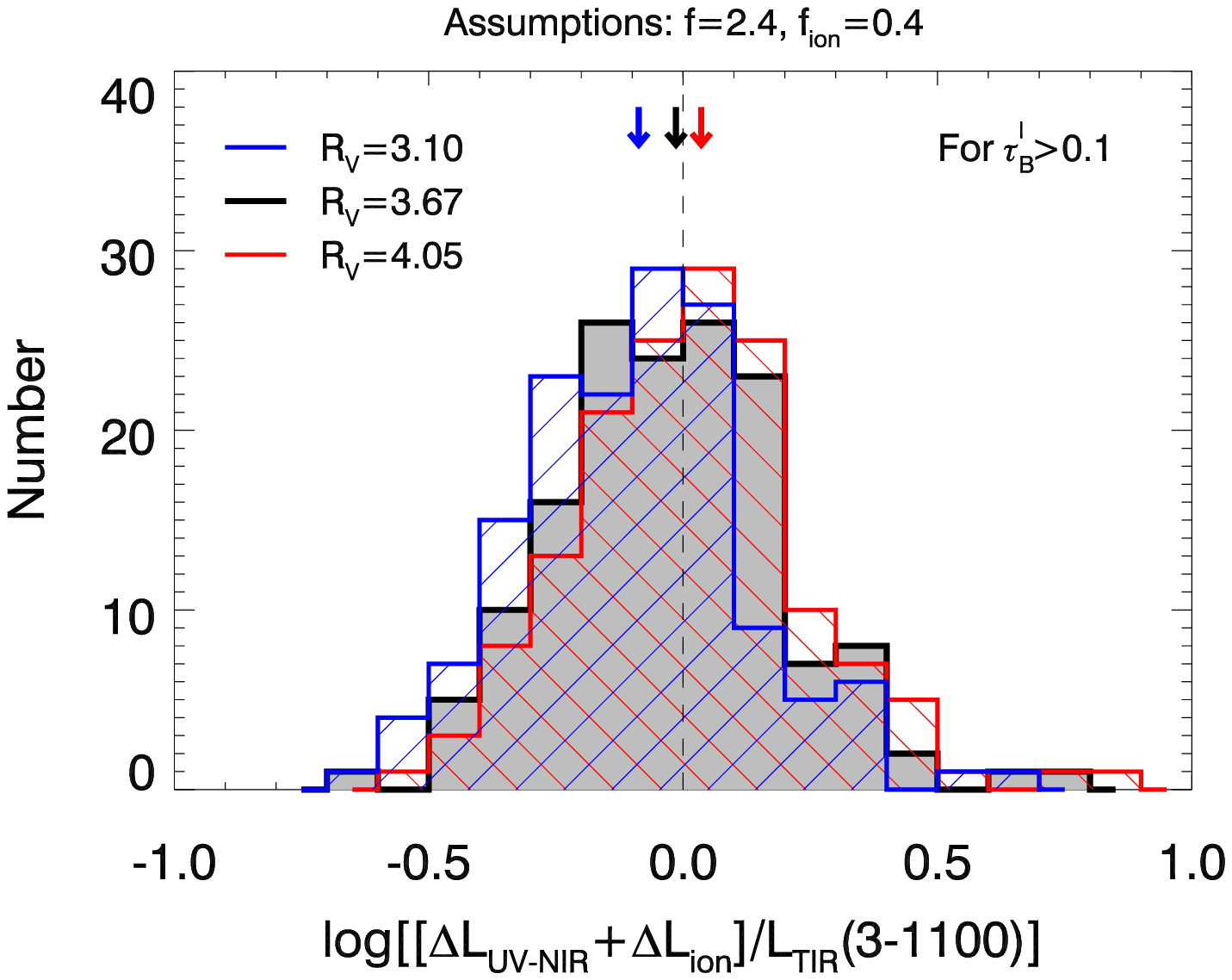} &
\includegraphics[width=2.2in]{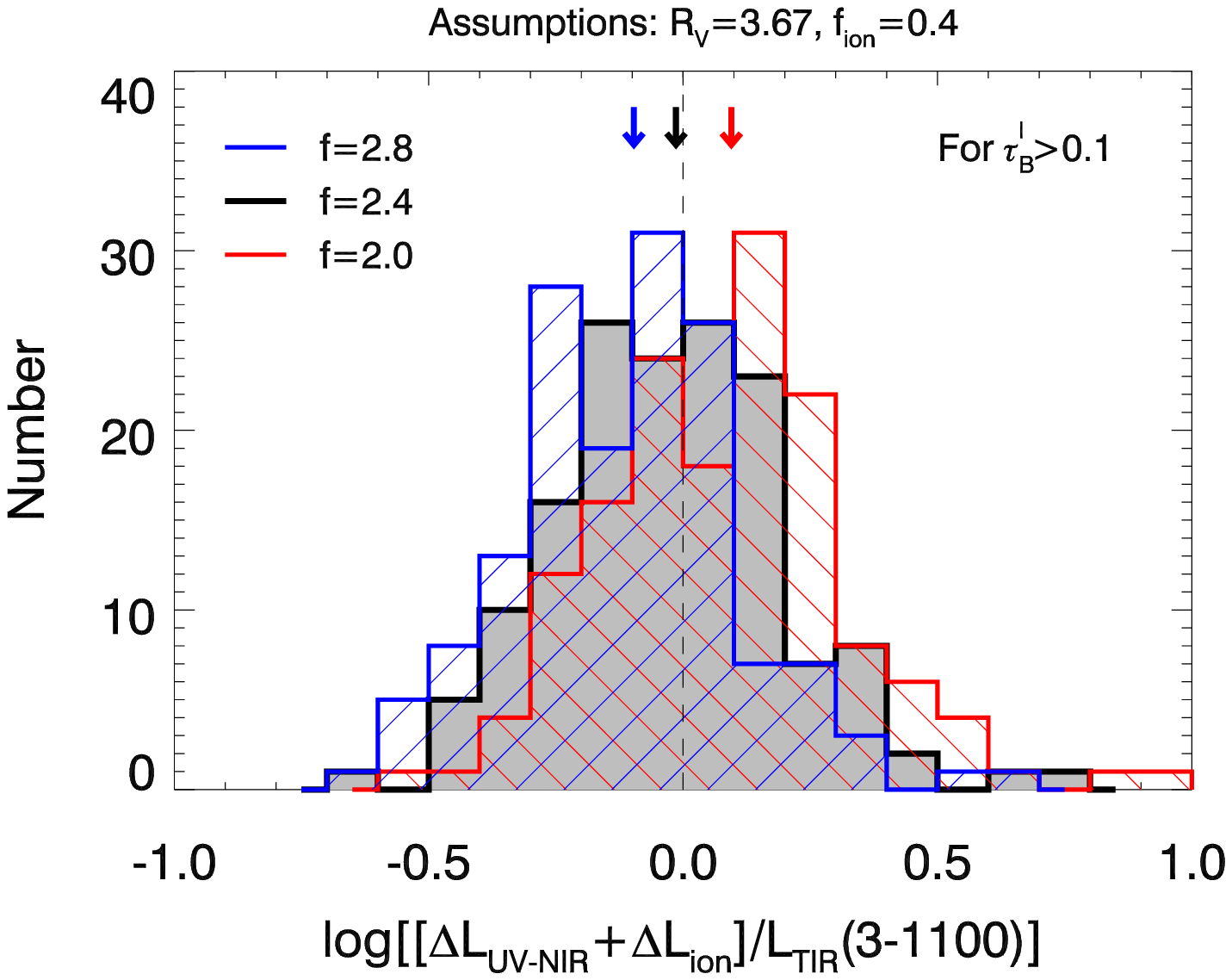}  &
\includegraphics[width=2.2in]{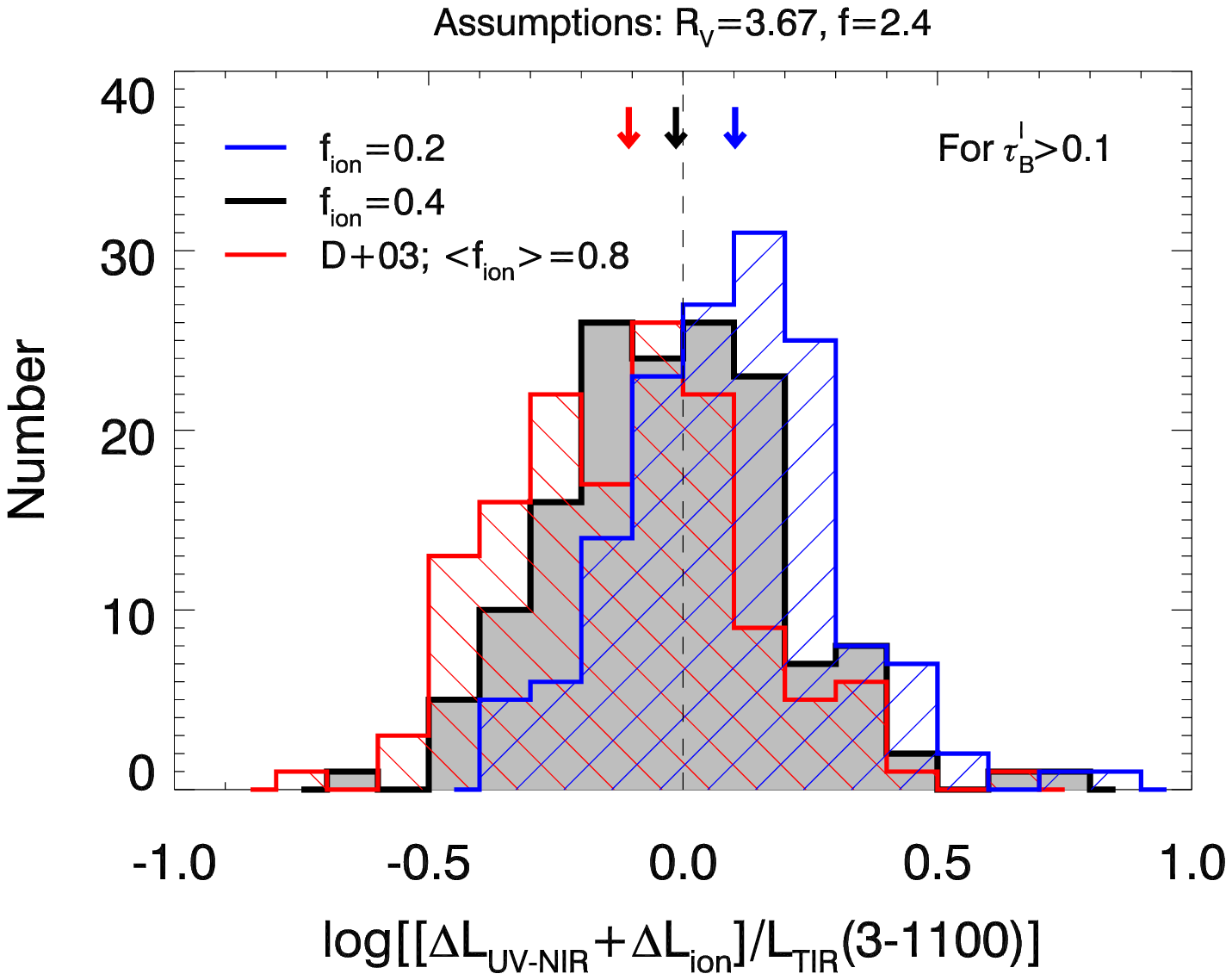} \\
\end{array}$
\end{center}
\caption{Energy balance comparison assuming different values for $R_V$, $f$, and $f_{\mathrm{ion}}$. We display only the sources with $\tau_B^l>0.1$.  The gray histogram is the same in each panel and corresponds to our fiducial assumptions (see Figure~\ref{fig:energy_ratio_compare}). A lower $R_V$ will result in lower total attenuation at all wavelengths, leading to lower energy ratios. The value of $f$ impacts the curve in an inverse manner, where a higher $f$ leads to lower ratios (due to smaller $E(B-V)_{\mathrm{star}}$). Larger values of $f_{\mathrm{ion}}$ imply a lower fraction of direct dust absorption for ionizing photons, leading to lower energy ratios. The red histogram for the $f_{\mathrm{ion}}$ comparison (\textit{Right}), makes use of the models of compact HII regions from \citet{dopita03}. The median value for each distribution is indicated by the arrow with matching color. Differences in the appropriate values for these variables in individual galaxies could give rise to some of the observed scatter.
 \label{fig:energy_ratio_vs_Rv_f}}
\end{figure*}

Under the scenario of energy balance described above, we expect that
\begin{equation}
\frac{\Delta L_{\mathrm{UV-NIR}} + \Delta L_{\mathrm{ion}}}{L_{\mathrm{TIR}}}=1 \,.
\end{equation}
The resulting energy ratio that we find for the GMACS sample is shown in Figure~\ref{fig:energy_ratio_compare}. It can be seen that as $\tau_B^l$ approaches zero, the estimated energy ratio also becomes very small, as is expected because it directly determines the value of the numerator. However, the fact that these galaxies still have far-IR emission is indicating that there is dust absorption despite the low observed Balmer optical depth, which may reflect issues with using this quantity as a probe of the dust content for the entire galaxy or with geometric variations between stars and dust. This could also be the result of optically thick star formation having a larger contribution to \ltir\ in seemingly low-$\tau_B^l$ sources  (discussed more in \S~\ref{intrinsic_UV}). For this reason, we only expect sources with higher values of $\tau_B^l$ to be in a regime where the energy balance scenario outlined in this section is applicable. We take this cutoff value to be $\tau_B^l>0.1$ because beyond this region the behavior of the energy ratio appears to show negligible trend with $\tau_B^l$. In this region, we find that with our assumptions the mean and median values are close to unity, with values of -0.03 and -0.01 in log-space, respectively, but with relatively large scatter (1$\sigma=0.23$). We examine if the variation in the energy balance ratio is related to differences in global galaxy properties (which may influence the attenuation curve) in Figure~\ref{fig:GMACS_param_vs_energy_ratio}. The properties examined are  $D_n4000$, total stellar mass, total sSFR, and gas-phase metallicity. The energy ratio does not appear to be correlated with these global properties, indicating that the scatter could be associated with other factors.

To better understand the large variation in the energy balance ratios, we will now explore impact of utilizing different assumptions for its calculation. The main assumptions being made are: (1) the normalization and shape of the dust attenuation curve, (2) the differential reddening between the ionized gas and the stellar continuum, and (3) the fraction of ionizing photons directly absorbed by dust. In order to investigate the influence of the attenuation curve on the inferred energy ratios, we can also examine the values if we assume the SB attenuation curve of \citet{calzetti00} ($R_V=4.05$; $f=2.659$). Interestingly, the resulting mean and median values are comparable to those stated above, with values of -0.02 and -0.01 in log-space, respectively (for $\tau_B^l>0.1$ ) and similar scatter (1$\sigma=0.23$). The reason for this similarity is that despite the slightly larger UV attenuation in the SB curve, the value of $f$ is higher for this curve and this leads to a decrease in the values of $E(B-V)_{\mathrm{star}}$ from Equation~(\ref{eq:EBV_star}) relative to  the SFG sample. If instead we adopt a starburst-like curve with our average value of $f$, then the energy ratios are larger with a median value of 0.06 in log-space. To further illustrate the effect of these variables, we demonstrate how the energy ratios inferred from different $R_V$ and $f$ values in Figure~\ref{fig:energy_ratio_vs_Rv_f}, in addition to showing the effects of varying $f_{\mathrm{ion}}$. For the latter, we also test using the models of compact HII regions from \citet{dopita03}, which parameterizes $f_{\mathrm{ion}}$ as a function of the ionization parameter and the metallicity (see their Equation~10). They consider a dust model composed of silicates, graphite or amorphous carbon, and polycyclic aromatic hydrocarbons (PAHs). For this analysis, we will use their model based on 20\% of carbon atoms locked up in PAHs. The ionization parameter is determined using the relationship from \citet[][their Equation~13]{kobulnicky&kewley04}, which uses the ratio of the [OIII] to [OII] lines and metallicity (12+log(O/H)) \citep{kewley&dopita02} and are obtained from the MPA/JHU measurements. The resulting values for $f_{\mathrm{ion}}$ based on the \citet{dopita03} model tend to be larger ($\langle f_{\mathrm{ion}}\rangle\sim0.8$), corresponding to lower direct dust absorption, than the galaxy-wide averages found in \citet{inoue01b}. It can be seen in Figure~\ref{fig:energy_ratio_vs_Rv_f} that differences in the appropriate values for these attenuation variables for individual galaxies could give rise to the observed scatter. 

We have carried the analysis in the present section to mainly provide a consistency check for our attenuation curve. Given that the large variation in energy balance ratios can be driven by several factors, we argue that our analysis does not provide evidence in favor of adopting a single attenuation curve on an individual galaxy basis. However, if the fiducial assumptions are reasonable average values for SFGs, then adopting the curve found in \S~\ref{normalization} will be a suitable choice for  broadly characterizing the attenuation in a large sample of galaxies.

\section{Discussion}
\subsection{The Intrinsic UV Slope, \texorpdfstring{$\beta_0$}{beta_0}}\label{intrinsic_UV}
An interesting results of section~\ref{energy_balance} is that the average value of the UV slope, $\beta$, for the ``intrinsic" SEDs, $F_{\mathrm{int}}(\lambda)$, after correcting by the observed reddening on the ionized gas ($\tau_B^l$) is $\beta_0=-1.61$, which corresponds to the zero-point of the $\beta$-$\tau_B^l$ relation \citep{battisti16}. This value is far redder than expected for an unreddened galaxy undergoing continuous star formation on timescales of $\sim$Gyr, for which $\beta_0\sim-2.1$ \citep{calzetti00}. Interestingly, if we apply attenuation corrections to the SEDs such that $\beta_0=-2.1$, then the inferred luminosity from dust absorption would drastically exceed the observed IR luminosity (i.e., $(\Delta L_{\mathrm{UV-NIR}} + \Delta L_{\mathrm{ion}})/L_{\mathrm{TIR}}>1$) for most of the sample. 

This apparent conflict may indicate that the young, massive stars responsible for most of the UV photons are heavily embedded in dust but that their optical depth is not reflected in the galaxy average, possibly due regions of low optical depth providing the majority of the flux density \citep{calzetti94}. This would support the scenario in which the dust content of galaxies has two components, one associated with short-lived dense clouds where massive stars form and another associated with the diffuse interstellar medium \citep[e.g.,][]{calzetti94,charlot&fall00,wild11}. Locally, it is found that the majority of \ltir\ is due to the cold, diffuse dust component, with only $\sim10\%$ on average being due to warm dust heated in photo-dissociation regions \citep{draine07}. Therefore, we can expect that if massive stars are being obscured to achieve $\beta_0=-1.61$, it may not lead to substantial increases in the observed \ltir. This can also explain why dust emission is seen for the $\tau_B^l\sim0$ cases, where presumably the warm dust component provides a larger fraction of \ltir. We will explore these possibilities in more detail in a subsequent paper. 

\subsection{Calculating the Total Attenuation}
A useful tool for recovering the intrinsic luminosity of a galaxy is the relationship between the total attenuation and the observed UV slope. We determine the average relationship for our sample of SFGs using the attenuation curve given by Equation~(\ref{eq:k_lambda}) and the $\beta$-$\tau_B^l$ relation from \citet{battisti16}, $\beta = (1.95\pm0.03)\times \tau_B^l -(1.61\pm0.01)$, with $\sigma_{\mathrm{int}}=0.44$. These give
\begin{equation}\label{eq:A_lambda}
A_\lambda = k(\lambda)[(0.232\pm0.029)\times(\beta+(1.61\pm0.44))]\,,
\end{equation}
where $A_\lambda = k(\lambda)E(B-V)_{\mathrm{star}}$ and we use Equation~(\ref{eq:EBV_star}) with $f=2.4\pm0.3$. Here we have also included the intrinsic scatter of $\beta$ into the uncertainty of $\beta_0=-1.61$. Assuming $k(1600)=8.76\pm0.4$, the total attenuation at 1600\AA\ is
\begin{equation}\label{eq:A_1600}
A_{1600} = (2.03\pm0.27)\beta+(3.27\pm0.99)\,.
\end{equation}
The dominant source of uncertainty in this relationship is the large intrinsic scatter in UV slope values for star-forming galaxies ($\sigma_{\mathrm{int}}=0.44$), and this severely limits its utility for directly indicating the total attenuation. This issue will be examined and addressed in a subsequent paper. Interestingly, Equation~(\ref{eq:A_1600}) is consistent with the results from \citet{casey14}, $A_{1600} = (2.04\pm0.08)\beta+(3.36\pm0.10)$, determined independently using the relationship between the infrared excess ($IRX\equiv L_\mathrm{IR}/L_\mathrm{UV}$) and the UV slope ($\beta$). For comparison, the relationship for local starburst galaxies is $A_{1600} = 2.31\beta+4.85$ \citep{calzetti00}, where this relation assumes $\beta_0=-2.1$.

\section{Conclusions}
Using a sample of $\sim$5500 local ($z\lesssim0.1$) star-forming galaxies we have characterized the average behavior of dust attenuation from the UV-to-NIR. This analysis utilizes the fact that the reddening of ionized gas, which is an excellent tracer of dust, is correlated with the reddening of the stellar continuum, albeit with noticeable scatter driven by variations in the underlying stellar population and/or differences in the star-dust geometry between different galaxies. 

We construct an average total-to-selective attenuation curve, $k(\lambda)$, from the selective attenuation by assuming that the curve should approach zero near $2.85\micron$. The resulting dust attenuation curve for local SFGs has a normalization at $V$-band of $R_V=3.67\substack{+0.44 \\ -0.35}$, which is slightly lower than the value seen for local starburst galaxies \citep[$R_V=4.05\pm0.80$,][]{calzetti00}. In addition, this curve is lower in the FUV than starburst galaxies by about 15\%, but becomes similar at longer wavelengths. Using a small sample of local galaxies with available far-IR data to independently constraint the dust attenuation, it is found that the total attenuated energy inferred from this curve shows general agreement with the observed dust emission (assuming energy balance), however with a significant level of individual variation (70\%). 

Both the derived shape and normalization of the attenuation curve carry significant uncertainties, partly due to the difficulty of breaking the degeneracy between stellar population age and attenuation; however, it could also indicate that the variation in the attenuation from galaxy to galaxy is large. We find minor variations in the average attenuation curve as a function of mean stellar population age ($D_n4000$), specific star formation rate, stellar mass, and metallicity, but we consider these trends to have low significance because of the large uncertainties and potential for systematic effects. We suggest that the attenuation curve given by Equation~(\ref{eq:k_lambda}) is well suited for applications to broadly characterize large samples of galaxies in the local Universe, but should be viewed with caution in its application to individual galaxies.

\section*{Acknowledgments}The authors thank the anonymous referee whose suggestions helped to clarify and improve the content of this work. AJB also thanks K. Grasha and V. Wild for comments that improved the clarity of this paper.

Part of this work has been supported by the National Aeronautics and Space Administration (NASA), via the Jet Propulsion Laboratory Euclid Project Office, as part of the ``Science Investigations as Members of the Euclid Consortium and Euclid Science Team'' program.

This work is based on observations made with the NASA Galaxy Evolution Explorer. \textit{GALEX} is operated for NASA by the California Institute of Technology under NASA contract NAS5-98034.
This work has made use of SDSS data. Funding for the SDSS and SDSS-II has been provided by the Alfred P. Sloan Foundation, the Participating Institutions, the National Science Foundation, the US Department of Energy, the National Aeronautics and Space Administration, the Japanese Monbukagakusho, the Max Planck Society and the Higher Education Funding Council for England. The SDSS website is http://www.sdss.org/. The SDSS is managed by the Astrophysical Research Consortium for the Participating Institutions.
This work has made use of data obtained with the United Kingdom Infrared Telescope. UKIRT is supported by NASA and operated under an agreement among the University of Hawaii, the University of Arizona, and Lockheed Martin Advanced Technology Center; operations are enabled through the cooperation of the East Asian Observatory. When the data reported here were acquired, UKIRT was operated by the Joint Astronomy Centre on behalf of the Science and Technology Facilities Council of the U.K.
This publication makes use of data products from the Two Micron All Sky Survey, which is a joint project of the University of Massachusetts and the Infrared Processing and Analysis Center/California Institute of Technology, funded by the NASA and the National Science Foundation.

\appendix
\section{Comparison of Near-IR Data}\label{NIR_compare}
\begin{figure*}[h]
\begin{center}$
\begin{array}{ccc}
\includegraphics[width=2.2in]{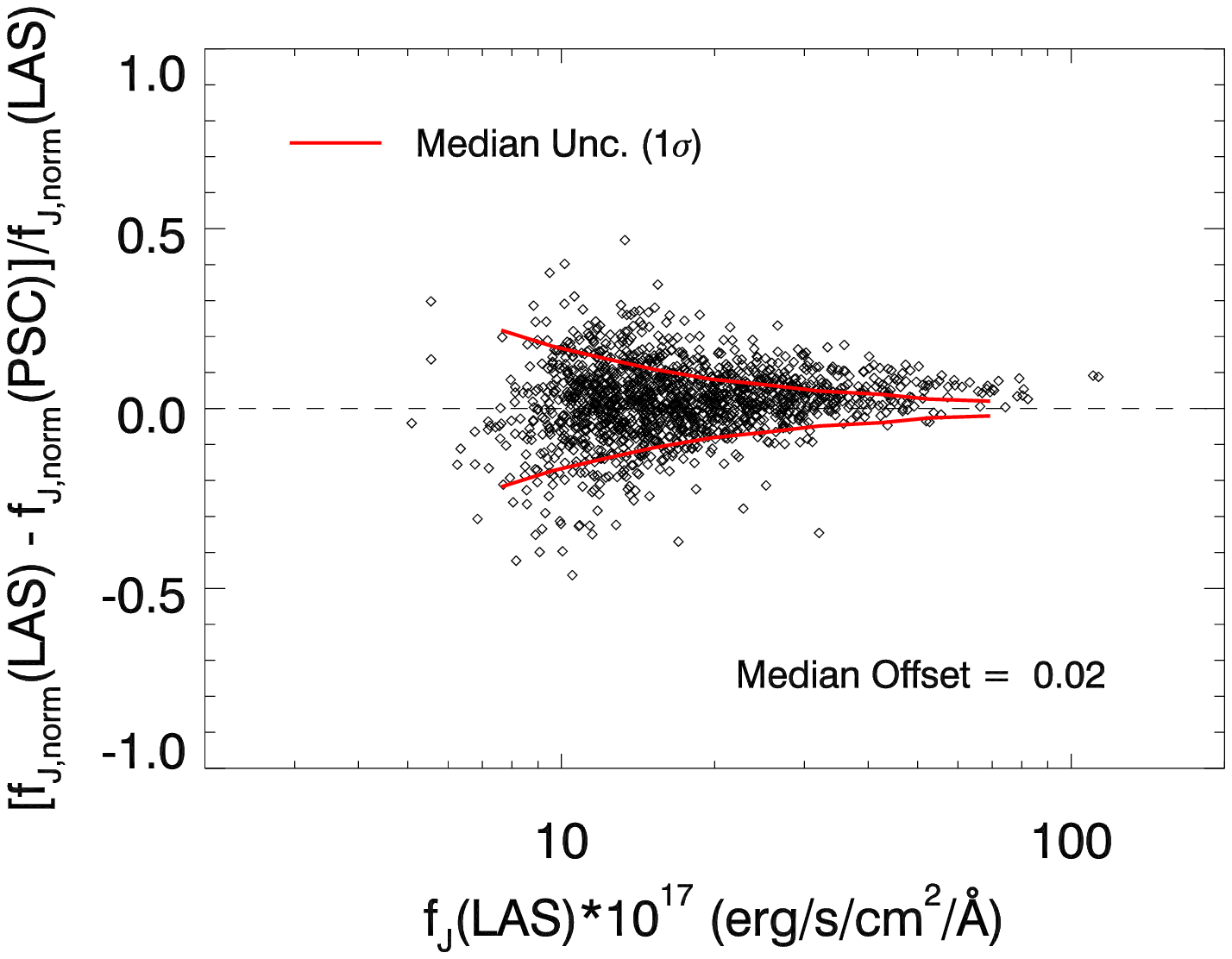} &
\includegraphics[width=2.2in]{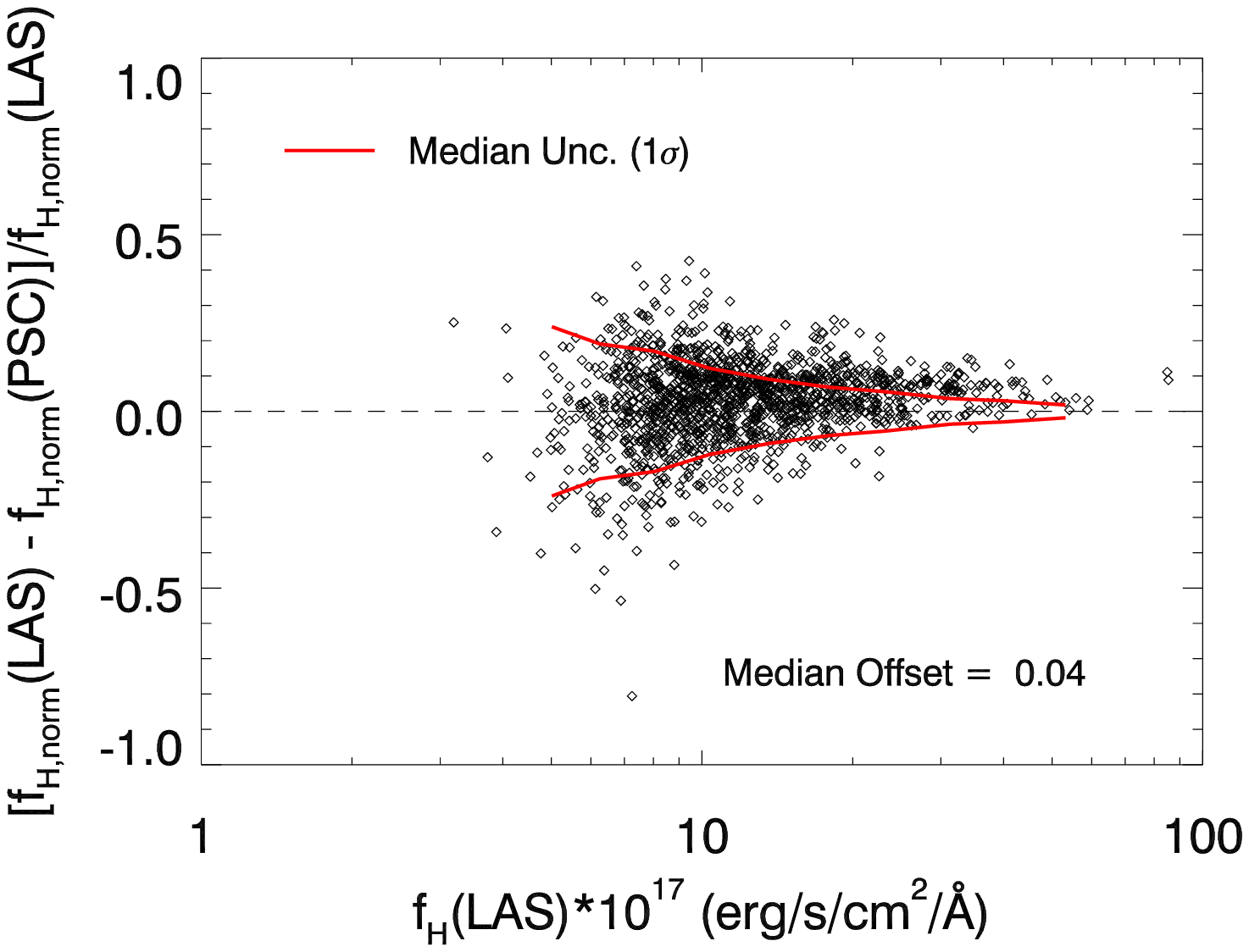} &
\includegraphics[width=2.2in]{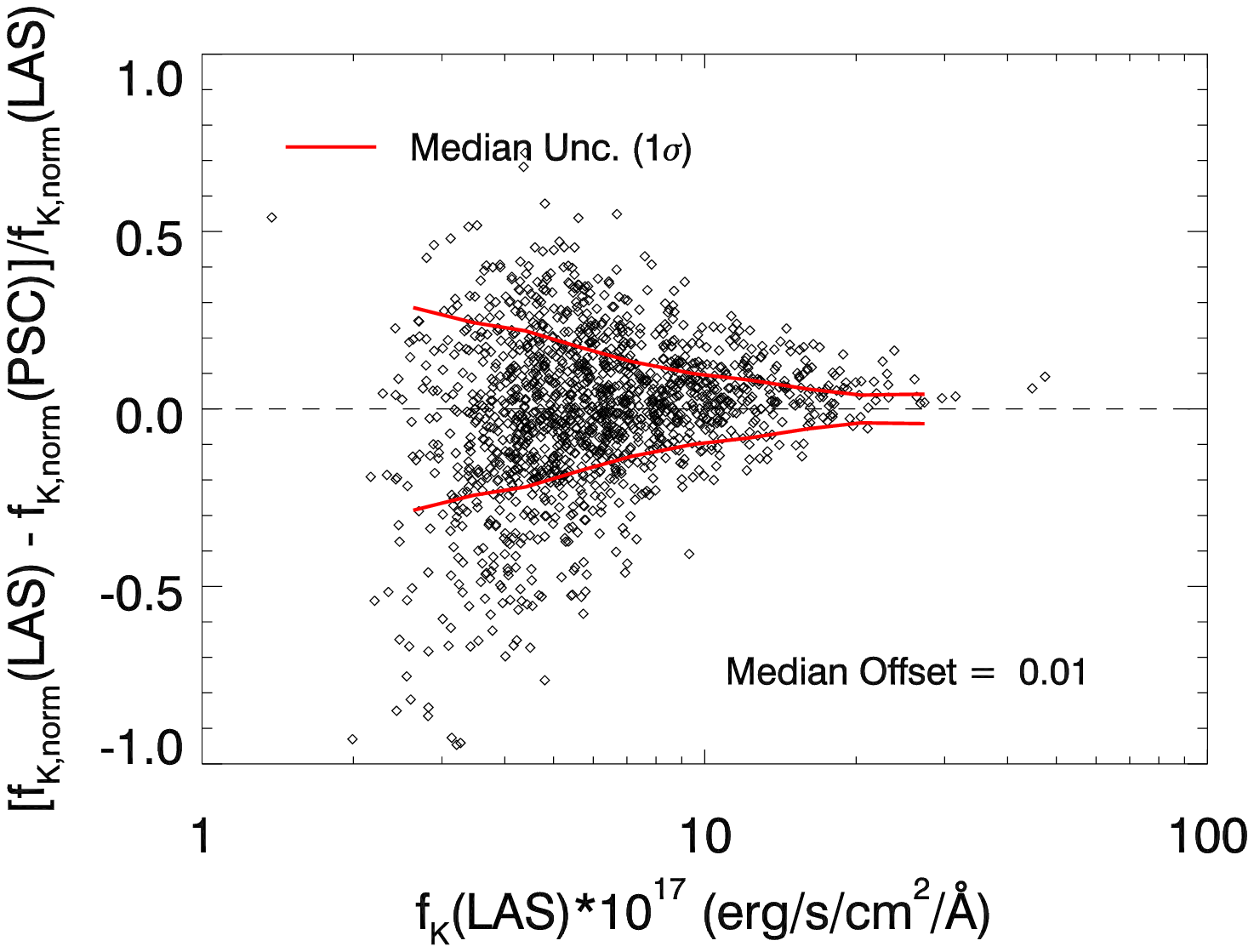} \\
\end{array}$
\end{center}
\caption{Comparison between sources in the UKIDSS LAS and the 2MASS PSC after normalizing the apertures according to their SDSS $z$-band values, where $f_{X,\mathrm{norm}}(\mathrm{LAS})=f_X(5.7")/f_z(5.7")$ and $f_{X,\mathrm{norm}}(\mathrm{PSC})=f_X(8.0")/f_z(8.0")$.
 \label{fig:compare_NIR}}
\end{figure*}

\begin{figure*}[h]
\begin{center}$
\begin{array}{ccc}
\includegraphics[width=2.2in]{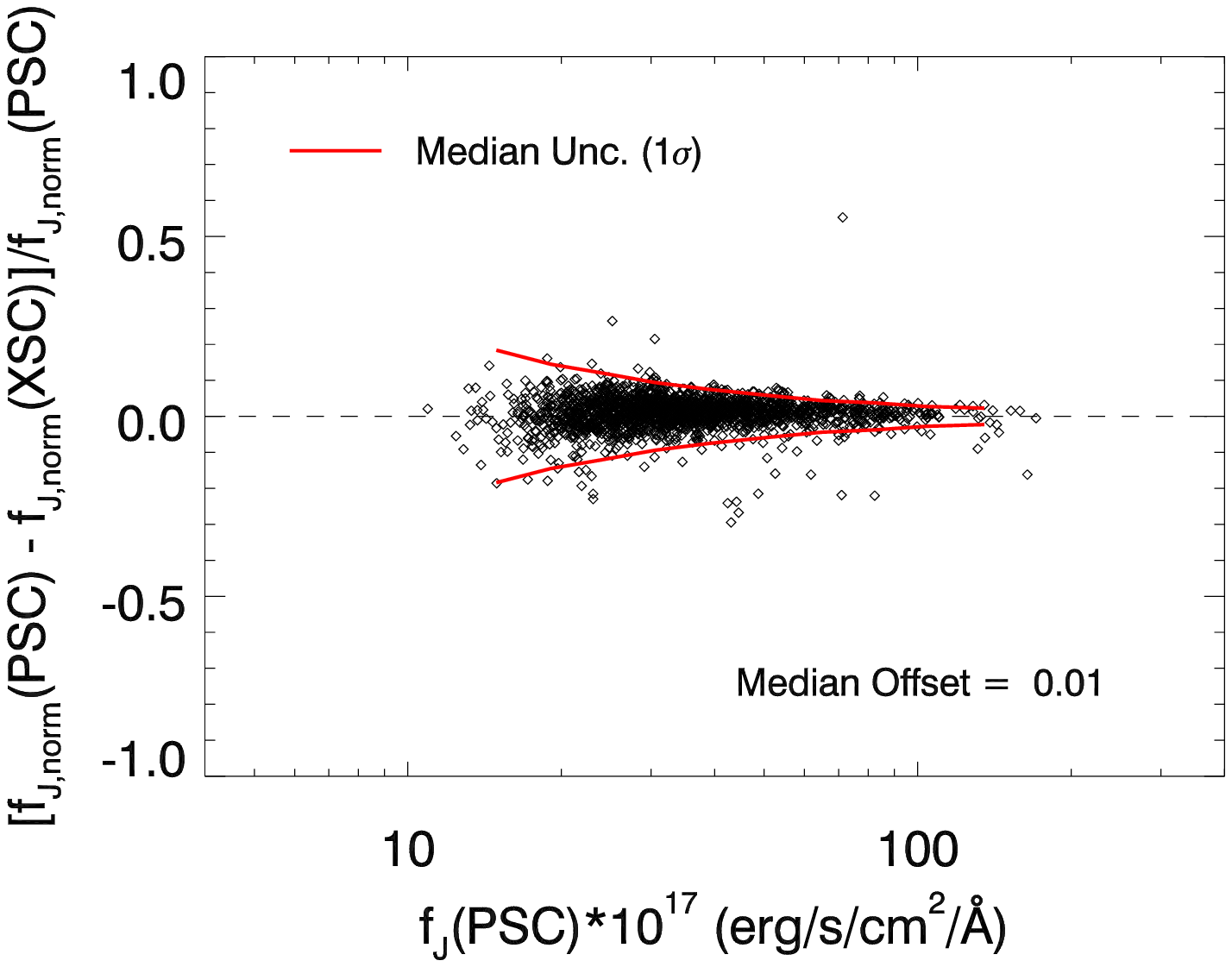} &
\includegraphics[width=2.2in]{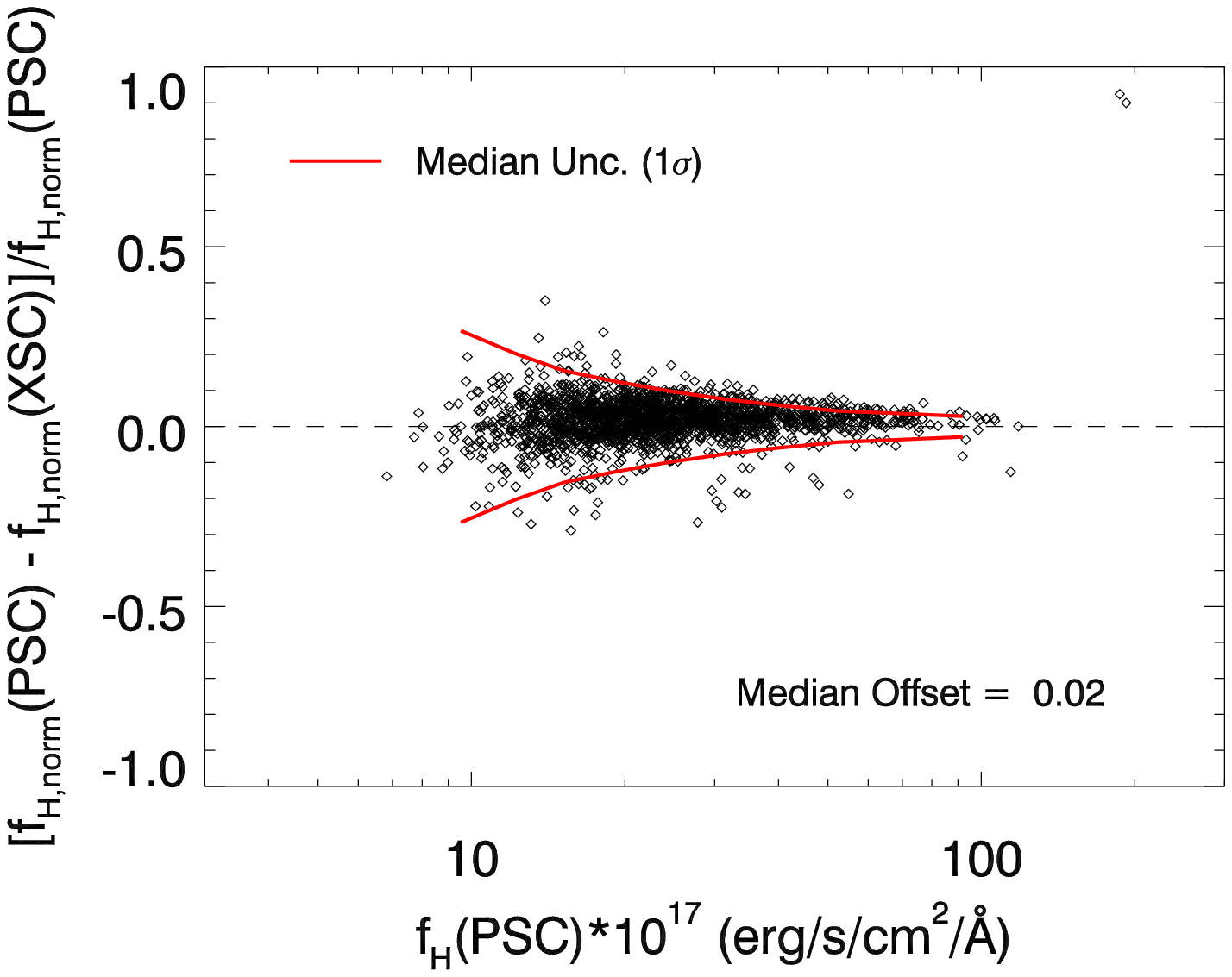} &
\includegraphics[width=2.2in]{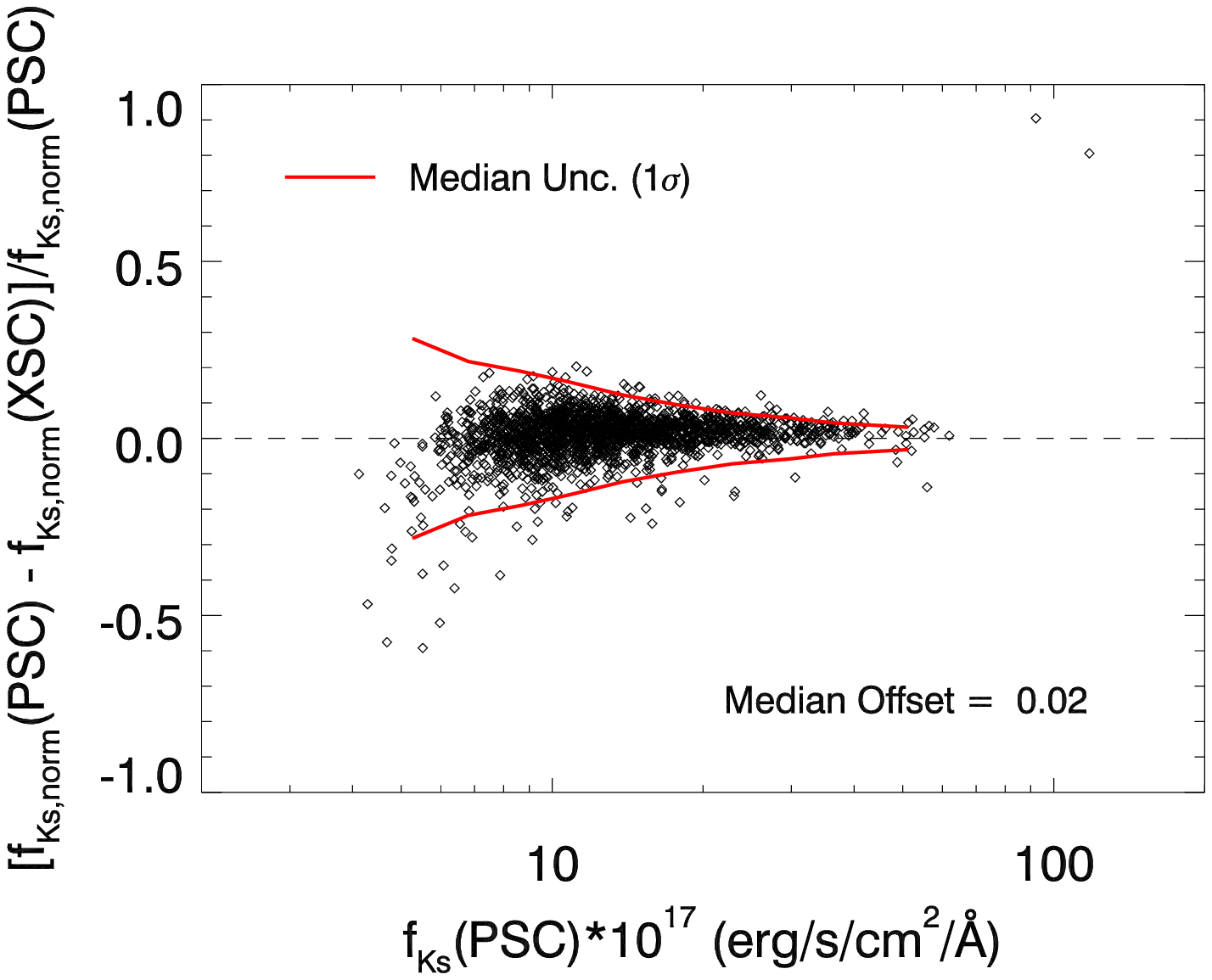} \\
\end{array}$
\end{center}
\caption{Comparison between sources in the 2MASS PSC and the 2MASS XSC after normalizing the apertures according to their SDSS $z$-band values, where $f_{X,\mathrm{norm}}(\mathrm{PSC})=f_X(8.0")/f_z(8.0")$ and $f_{X,\mathrm{norm}}(\mathrm{XSC})=f_X(10")/f_z(10")$.
 \label{fig:compare_2MASS}}
\end{figure*}

In this section we show the comparison between the different aperture-normalized NIR datasets examined in this study. We normalize the NIR data using the SDSS $z-band$ value aperture photometry at 4.5, 5.7, and $8.0\arcsec$. We normalize the UKIDSS and 2MASS photometry such that the ratio between the NIR and the SDSS $z$-band flux density at the catalog aperture is preserved for $4.5\arcsec$ (e.g., UKIDSS: $f_J(4.5\arcsec)=f_z(4.5\arcsec)[f_J(5.7\arcsec)/f_z(5.7\arcsec)]$; 2MASS: $f_J(4.5\arcsec)=f_z(4.5\arcsec)[f_J(8\arcsec)/f_z(8\arcsec)]$). The comparison between UKIDSS LAS and 2MASS PSC photometry after applying these normalizations are shown in Figure~\ref{fig:compare_NIR}. There is good agreement between the data, with a majority of cases lying within the median 1$\sigma$ uncertainty (red line). We also show a comparison between the overlapping sources in the 2MASS PSC and XSC in Figure~\ref{fig:compare_2MASS}, which also demonstrate good agreement.

\end{document}